\documentclass[aps,prc,groupedaddress,showpacs,letterpaper,floatfix]{revtex4}
\usepackage{dcolumn}
\usepackage{xspace}
\usepackage{amssymb}
\usepackage{amsmath}
\usepackage{graphicx}
\usepackage{colordvi}

\newcommand{\nuc}[2]{\ensuremath{\rm{^{#1}}#2}}                       
\newcommand*{\reaction}[6]{\nuc{#1}{#2}(#3,#4)\/\nuc{#5}{#6}}
\newcommand*{\degree}{\ensuremath{^{\circ}}\xspace}
\newcommand*{\ayy}{\ensuremath{A_{yy}}\xspace}

\newcommand*{\ayhe}{\ensuremath{A_{0y}}\xspace}
\newcommand*{\ay}{\ensuremath{A_{y}}\xspace}

\newcommand*{\dsig}{\ensuremath{\sigma(\theta)}\xspace}
\newcommand*{\Ecm}{\ensuremath{E_{\mathrm{c.m.}}\xspace}}
\newcommand*{\Tcm}{\ensuremath{T_{\mathrm{c.m.}}\xspace}}

\newcommand*{\Ep}{\ensuremath{E_{p}}\xspace}
\newcommand*{\Ehe}{\ensuremath{E_{\mathrm{\nuc{3}{He}}}\xspace}}
\newcommand*{\tlab}{\ensuremath{\theta_{\mathrm{lab}}}\xspace}
\newcommand*{\tcm}{\ensuremath{\theta_{\mathrm{c.m.}}}\xspace}

\newcommand*{\NN}{\ensuremath{NN}\xspace}
\newcommand*{\Nd}{\mbox{$N$-$d$}\xspace}
\newcommand*{\nd}{\mbox{$n$-$d$}\xspace}
\newcommand*{\pd}{\mbox{$p\,$-$d$}\xspace}

\newcommand*{\phe}{\textit{p}-\nuc{3}{He}\xspace}
\newcommand*{\nt}{\textit{n}-\nuc{3}{H}\xspace}

\newcommand{\by}{\ensuremath{\times}\xspace}

\def\het{{{}^3{\rm He}}}

\def\jac{x}

\begin{document}



\title{Proton - $^{3}\mathrm{He}$ Elastic Scattering at Low Energies}



\author{B. M. Fisher}
	\email[Electronic address: ]{brian.fisher@nist.gov}
	\altaffiliation[Present address: ]{Tulane University, New Orleans, LA 70118 USA;
                    and National Institute of Standards and Technology, Gaithersburg, MD 20899 USA}
\author{C. R. Brune}
	\altaffiliation[Present address: ]{Department of Physics and Astronomy, Ohio University, Athens, OH 45701 USA}
\author{H. J. Karwowski}
\author{D. S. Leonard}
	\altaffiliation[Present address: ]{Department of Physics and Astronomy, University of Alabama, Tuscaloosa, AL 35487 USA}
\author{E. J. Ludwig}
\affiliation{Department of Physics and Astronomy, University of North Carolina at Chapel Hill, Chapel Hill, North Carolina 27599-3255 USA\\
	and\\
	Triangle Universities Nuclear Laboratory, Durham, North Carolina 27708-0308 USA}

\author{T. C. Black}
\affiliation{Department of Physics and Physical Oceanography, University of North Carolina at Wilmington, Wilmington, North Carolina 28403 USA}

\author{M. Viviani}
\author{A. Kievsky}
\author{S. Rosati}
\affiliation{Istituto Nazionale di Fisica Nucleare, Sezione di Pisa,\\
             and\\
             Department of Physics, University of Pisa, I-56100 Pisa, Italy}


\date{\today}

\begin{abstract}

We present new accurate measurements of the differential cross section $\sigma(\theta)$ and the proton analyzing power $A_{y}$ for proton-$^{3}\mathrm{He}$ elastic scattering at various energies. A supersonic gas jet target has been employed to obtain these low energy cross section measurements.  The $\sigma(\theta)$ distributions have been measured at $E_{p}$ = 0.99, 1.59, 2.24, 3.11, and 4.02 MeV.  Full angular distributions of $A_{y}$ have been measured at  $E_{p}$ = 1.60, 2.25, 3.13, and 4.05 MeV. This set of high-precision data is compared to four-body variational calculations employing realistic nucleon-nucleon ($NN$) and three-nucleon ($3N$) interactions. For the unpolarized cross section the agreement between the  theoretical calculation and data is good when a $3N$ potential is used. The comparison between the calculated and measured proton analyzing powers reveals discrepancies of approximately 50\% at the maximum of each distribution.  This is analogous to the existing ``$A_{y}$ Puzzle'' known for the past 20 years in nucleon-deuteron elastic scattering.

\end{abstract}




\maketitle


\section{Introduction}
\label{sec:intro}

The study of the four nucleon ($4N$) system is interesting from a number of different perspectives. First of all, many reactions involving four nucleons, like \reaction{2}{H}{$d$}{$p$}{3}{H}, \reaction{2}{H}{$d$}{$n$}{3}{He}, or $p + \nuc{3}{He} \rightarrow \nuc{4}{He} + \nu_{e} + e^{+}$ (the $hep$ process), are of extreme astrophysical interest, as they play important roles in solar models or in big-bang nucleosynthesis (BBN).  The $hep$ process, for instance, is the source of the highest energy neutrinos from the Sun.  Moreover, $4N$ systems have become increasingly important as testing grounds for models of the nuclear force.  While not the nuclear-structure ``imbroglio'' of heavy nuclei, the $A = 4$ system is the simplest system that presents the complexity --- thresholds and resonances --- that characterize nuclear systems, and therefore is a very good testing ground of modern few-body techniques~\cite{car01}.  Similarly, since the $4N$ bound state is (to a very good approximation) a $(J^{\pi}, T) = (0^{+}, 0)$ state, it is a good ``laboratory'' for the study of the strange quark components of the nucleon via parity-violating electron-scattering experiments~\cite{bec01}.

The theoretical description of $A=4$ systems still constitutes a challenging problem from the standpoint of nuclear few-body theory.  Only recently, with the near-constant increase in computing power and the development of new numerical methods, has the study of the $\alpha$-particle bound state reached a satisfactory level of accuracy; the $4N$ bound-state has been calculated to a few tenths of keV \cite{kam01, nog02, viv05}. The study of $4N$ scattering states, on the other hand, is less satisfactorily developed.  The same increases in computational power, however, have opened the possibility for accurate calculations of the $4N$ observables using realistic models for nucleon-nucleon (\NN) and three-nucleon ($3N$) forces.  These calculations have been performed mainly by means of the Faddeev-Yakubovsky (FY) approach~\cite{cie98, fon99, laz04} and the Kohn variational principle~\cite{viv98, viv01, hof99, pfi01, rei03}. In spite of this progress, some disagreements still exist between theoretical groups, as for example, in the calculation of the \nt total cross section in the peak region (around $\Ecm = 3$ MeV)~\cite{laz05}.

To make matters worse, even in cases where the theoretical calculations agree, they are often strongly at variance with the experimental data. For example, the proton analyzing power \ay in \phe elastic scattering is underestimated by theory at the peak of the angular distribution by about 40\% \cite{viv01}. Other disagreements are discussed in Refs. \cite{geo03, fon99, pfi01, car01, laz04}.

The existing data in the literature for 4$N$ scattering are of lesser quality, when compared with the excellent and abundant data that exists for the \NN and $3N$ (\Nd scattering) systems. The intensely studied \nuc{4}{He} system is (unfortunately) very difficult to describe theoretically due to the presence of the bound state and many higher-energy resonances~\cite{til92}. This leads us to investigate \phe scattering states.

\subsection{$n$-\nuc{3}{H} Elastic Scattering}
\label{sec:intro:nt}

This paper will focus on \phe elastic scattering at low energies, which is simpler than the \nuc{4}{He} system to investigate theoretically.  However, the situation for the closely related \nt elastic scattering case is worth briefly discussing first.  The quantities of interest in \nt zero-energy scattering are the zero-energy total cross section $\sigma_{T}(0)$ and the coherent scattering length $a_{c}$.  Experimentally, only the total cross section $\sigma_{T}$ has been measured with high precision for a large range of energies.    The extrapolation of the measured $\sigma_{T}$ to zero energy is straightforward and the value obtained is $\sigma_{T}(0) = 1.70 \pm 0.03$ b~\cite{phi80}.  The coherent scattering length has been measured by neutron-interferometry techniques.  The most recent value reported in the literature is $a_{c} = 3.59 \pm 0.02$ fm~\cite{rau85}.  An additional estimation of $a_{c} = 3.607 \pm 0.017$ fm has been obtained from \phe data by using an approximate Coulomb-corrected R-matrix theory~\cite{hal90}. These values should be compared with those obtained theoretically: $\sigma_{T}(0) = 1.73$ b and $a_{c} = 3.71$ fm~\cite{viv98}.

As already mentioned, at higher energies (specifically in the ``peak'' region) there exist sizable discrepancies between different theoretical predictions.  The $\sigma_{T}$ calculations by Fonseca \cite{fon99} and Pfitzinger, Hofmann, and Hale~\cite{pfi01} are in  good agreement with the experimental data, while the calculations by the Grenoble~\cite{cie99, laz04, laz05} and Pisa groups~\cite{laz05} are well below the data.  The origin of this disagreement is still not clear.

\subsection{\phe Elastic Scattering}
\label{sec:intro:phe}

Let us consider now the situation for \phe elastic scattering.  The zero-energy quantities for this case are more difficult to evaluate. Approximate values of the triplet and singlet scattering lengths have been determined from effective range extrapolations \cite{all93b} to zero energy of data taken mostly above 1 MeV, and therefore suffer large uncertainties~\cite{viv01}. This problem has been reconsidered recently by George and Knutson~\cite{geo03}; this new phase-shift analysis (PSA) gave two possible sets of scattering-length values, both of which are at variance with the theoretical estimates (see the discussion in Ref.~\cite{geo03}.)

The world database of existing \phe scattering data can be divided in three energy regions.  There is a set of cross section and proton analyzing power \ay measurements  at very low energies, from $\Ep = 0.3$ to $1.0$ MeV~\cite{ber80}.  Another energy region is for $\Ep = 1.0$ to $4.0$ MeV which includes the recent \ay measurements at $\Ep = 1.60$ and $2.25$  MeV \cite{viv01}.  However, these \ay data are not very precise.  The cross-section data in this region are similarly imprecise and sparse in number.  They are also very old, being the very first published cross-section data for \phe elastic scattering found in the literature~\cite{fam54}.   The third group of measurements (for $\Ep > 4.0$ MeV) includes differential cross sections, proton and \nuc{3}{He} analyzing power measurements, and spin correlation coefficients of good precision. References for all these measurements can be found in Ref. \cite{all93b}.

The calculations performed so far for \phe scattering have shown a glaring discrepancy between theory and experiment in the proton analyzing power~\cite{viv01,rei03}. This discrepancy is very similar to the well known ``\ay  Puzzle''  in \Nd scattering. This is a fairly old problem, already reported almost 30 years ago~\cite{koi87, wit89} in the case of \nd and later confirmed also in the \pd case~\cite{kie96,woo02}.  All \Nd theoretical calculations based on realistic \NN potentials (even including $3N$ forces) underestimate the measured nucleon vector analyzing power \ay by about $20-30$ \%. The same problem also occurs for the vector analyzing power of the deuteron ${\mathrm{i}}T_{11}$~\cite{kie96,woo02}, while the tensor analyzing powers $T_{20}$, $T_{21}$ and $T_{22}$ are reasonably well described~\cite{glo96,woo02}.  The inclusion of standard models of the $3N$ force have little effect on calculations of these observables.  To solve this puzzle, speculations about the deficiency of the \NN potentials in ${}^3P_j$ waves (where the spectroscopic notation $^{2S+1}L_J$ has been adopted) have been suggested.  Looking at \NN scattering data only, this possibility does not seem to be ruled out~\cite{tor98, tor99}.  However, after taking theoretical constraints into account~\cite{hub98}, this solution was considered unlikely, and has been confirmed by recent calculations which show that this puzzle is not solved even when new \NN potentials derived from effective field theory are used~\cite{laz04, ent02}.

Consequently, attention has focused on exotic $3N$ force terms not contemplated so far.  The best solution seems to be given by the inclusion of a spin-orbit $3N$ force~\cite{kie99}; such a force has a negligible effect on the observables already well reproduced by standard $3N$ potential models (such as the binding energies, \Nd unpolarized cross sections and tensor analyzing powers) but it is very effective for solving --- or at least for noticeably reducing --- the discrepancy. Other explanations have been proposed in Refs.~\onlinecite{ishi99, can01, sam00, friar01}.  It should be noted that the current understanding of the 3N interaction is rather poor~\cite{friar99, epe02}, and new terms in the $3N$ interactions derived from chiral perturbation theory have very recently been proposed~\cite{epe05, friar05}.  These new models have to be tested primarily in the $3N$ system, but the $4N$ system could also play an important role.  In fact, $3N$ force effects are expected to be sizeable in the $4N$ system \cite{car98}; in the \Nd scattering system $3N$ force effects are small~\cite{glo96} and masked in part by the contribution of the Coulomb potential~\cite{kie01, kie04}.  Moreover, the \Nd system is essentially a pure isospin $T = 1/2$ state.  Tests of the $T = 3/2$ channel in any $3N$ force can only be satisfactorily performed in a $4N$ system such as \phe.

Seeking to explore potential three-nucleon force effects in a different system, and to investigate this new $A=4$ \ay Puzzle, a series of proton- \nuc{3}{He} elastic scattering measurements have been made.  Angular distributions of the differential cross section \dsig and proton analyzing power \ay have been measured at several energies below 5 MeV; the analyzing power experiments were performed with a gas-cell target and the \dsig measurements were performed using a supersonic gas-jet target.  The nominal proton energy \Ep at which each experiment was performed is listed in Table \ref{table:exp:energies}.  These energies were chosen to maximize the improvement in the low energy database, and because theoretical calculations are tractable in this region.

In addition, we present new theoretical calculations for the \phe system.  They are performed via an expansion of the wave functions of the scattering states in terms of a hyperspherical harmonic (HH) basis and using the complex form of the Kohn variational principle~\cite{sch88, kie97a}.  These new calculations reach a much higher degree of accuracy than those performed previously using the correlated hyperspherical harmonic (CHH) functions~\cite{viv01}. In this paper we present calculations based on the Argonne $v_{18}$ (AV18)~\cite{wir95} \NN potential, which represents the \NN interaction in its full richness, with short-range repulsion, tensor and other non-central components and charge symmetry breaking terms.  The calculations are performed without and with the inclusion of the Urbana IX (UIX) $3N$ force~\cite{pud97}.

The remainder of this paper is organized as follows: in Section \ref{sec:theory}, a brief description of the HH technique is reported; in Section \ref{sec:exp}, the experimental setup and methods are discussed; the cross section \dsig and proton analyzing power \ay measurements are compared with the theoretical results in Section \ref{sec:compare}; finally, Section \ref{sec:conclusions} is devoted to the conclusions and broader context of the present work.

\section{The HH Technique for Scattering States}
\label{sec:theory}

The wave function $\Psi_{1+3}^{L S J J_z\pi}$ describing a \phe scattering state with incoming orbital angular momentum $L$ and channel spin $S$ ($S=0, 1$), total angular momentum $J$, and parity $\pi = (-)^L$ can be written as
\begin{equation}
    \Psi_{1+3}^{LSJJ_z\pi}=\Psi_C^{LSJJ_z\pi}+\Psi_A^{LSJJ_z\pi} \ ,
    \label{eq:psica}
\end{equation}
where $\Psi_C^{LSJJ_x\pi}$ vanishes in the limit of large intercluster separations, and hence describes the system in the region where the particles are close to each other and their mutual interactions are strong. On the other hand, $\Psi_A^{LSJJ_z\pi}$ describes the relative motion of the two clusters in the asymptotic regions, where the \phe interaction is negligible. In the asymptotic region the wave function $\Psi_{1+3}^{LSJJ_z\pi}$ reduces to $\Psi_{A}^{LSJJ_z\pi}$, which must therefore be the appropriate asymptotic solution of the Schr\"{o}dinger equation. $\Psi_{A}^{LSJJ_z\pi}$ can be decomposed as a linear combination of the following functions
\begin{equation}
  \Omega_{LSJJ_z}^\pm = \frac{1}{\sqrt{4}} \sum_{i=1}^4
  \Bigl [ [s_i \otimes \phi_3(jkl)]_{S} \otimes 
  Y_{L}(\hat{\bf y}_i) \Bigr ]_{JJ_z}
  \left ( f_L(y_i) {\frac{G_{L}(\eta_C,qy_i)}{q y_i}}
          \pm \text{i} {\frac{F_L(\eta_C,qy_i)}{q y_i}} \right ) \ ,
  \label{eq:psiom}
\end{equation}
where $\mathbf{y}_i$ is the distance vector between the proton (particle $i$) and \nuc{3}{He} (particles $jkl$), $q$ is the magnitude of the relative momentum between the two clusters, $s_i$ the spin state of particle $i$, and $\phi_3$ is the $\het$ wave function.  The total kinetic energy $\Tcm$ in the center of mass (c.m.) system and the proton kinetic energy \Ep in the laboratory system are
\begin{equation}
	\Tcm = \frac{q^2}{2\mu}\ , \qquad
	\Ep = {\frac{4}{3}} \Tcm\ ,\label{eq:energy}
\end{equation}
where $\mu = (3/4)M_N$ is the reduced mass ($M_N$ is the nucleon mass.)  Moreover, $F_L$ and $G_L$ are the regular and irregular Coulomb functions, respectively, with $\eta_C = 2\mu e^2/q$. The function $f_L(y_i)=[1-\exp(-\gamma y_i)]^{2 L+1}$ in~(\ref{eq:psiom}) has been introduced to regularize  $G_L$ at small $y_i$, and $f_L(y_i) \rightarrow 1$ as $y_i$ becomes large, thus not affecting the asymptotic behavior of $\Psi_{1+3}^{LSJJ_z\pi}$.  Note that for large values of $qy_i$,
\begin{equation}
  f_L(y_i) G_{L}(\eta_C,qy_i)\pm {\rm i} F_L(\eta_C,qy_i) \rightarrow
   \exp\Bigl[\pm {\rm i} (q y_i-L\pi/2-\eta_C\ln(2qy_i)+\sigma_L) \Bigr]\ ,
\end{equation}
and therefore, $\Omega_{LSJJ_z}^+$  ($\Omega_{LSJJ_z}^-$) describes in the asymptotic regions an outgoing (incoming) \phe relative motion. Finally,
\begin{equation}
  \Psi_A^{LSJJ_z\pi}= \sum_{L^\prime S^\prime}
 \bigg[\delta_{L L^\prime} \delta_{S S^\prime} \Omega_{LSJJ_z}^-
  - {\cal S}^J_{LS,L^\prime S^\prime}(q)
     \Omega_{L^\prime S^\prime JJ_z}^+ \bigg] \ ,
  \label{eq:psia}
\end{equation}
where the parameters ${\cal S}^J_{LS,L^\prime S^\prime}(q)$ are the $S$-matrix elements which determine phase-shifts and (for coupled channels) mixing angles at the energy \Tcm.  Of course, the sum over $L^\prime$ and $S^\prime$ is over  all values compatible with a given $J$ and parity. In particular, the sum over $L^\prime$ is limited to include either even or odd values such that $(-1)^{L^\prime}=\pi$.

The ``core'' wave function $\Psi^{LSJJ_z\pi}_C$ is expanded using the HH basis. For four equal mass particles, a suitable choice of the Jacobi vectors is
\begin{eqnarray}
    {\bf x}_{1p}& = & \sqrt{\frac{3}{2}} 
    \left ({\bf r}_m - \frac{ {\bf r}_i+{\bf r}_j +{\bf r}_k}{3} \right )\ , \nonumber\\
    {\bf x}_{2p} & = & \sqrt{\frac{4}{3}}
    \left ({\bf r}_k-  \frac{ {\bf r}_i+{\bf r}_j}{2} \right )\ , \label{eq:JcbV}\\
    {\bf x}_{3p} & =& {\bf r}_j-{\bf r}_i\ , \nonumber
\end{eqnarray}
where $p$ specifies a given permutation corresponding to the order $i$, $j$, $k$ and $m$ of the particles. By definition, the permutation $p=1$ is chosen to correspond to the order $1$, $2$, $3$ and $4$.  For a given choice of the Jacobi vectors, the hyperspherical coordinates are given by the so-called hyperradius $\rho$, defined by
\begin{equation}
   \rho=\sqrt{x_{1p}^2+x_{2p}^2+x_{3p}^2}\ ,\qquad ({\rm independent\
    of\ }p)\ ,
    \label{eq:rho}
\end{equation}
and by a set of angular variables which in the Zernike and Brinkman~\cite{zerni, F83} representation are i) the polar angles $\hat \jac_{ip}\equiv (\theta_{ip},\phi_{ip})$  of each Jacobi vector, and ii) the two additional ``hyperspherical'' angles $\phi_{2p}$ and $\phi_{3p}$ defined as
\begin{equation}
    \cos\phi_{2p} = \frac{ x_{2p} }{\sqrt{x_{1p}^2+x_{2p}^2}}\ ,
    \qquad
    \cos\phi_{3p} = \frac{ x_{3p} }{\sqrt{x_{1p}^2+x_{2p}^2+x_{3p}^2}}\ ,
     \label{eq:phi}
\end{equation}
where the $x_{ip}$ are the moduli of the Jacobi vector ${\bf{x}_i}$. The set of angular variables $\hat x_{1p}, \hat x_{2p}, \hat x_{3p}, \phi_{2p}$, and $\phi_{3p}$ is denoted hereafter as $\Omega_p$.  A generic HH function is written as
\begin{equation}
  {\cal Y}^{K,\Lambda, M}_{\ell_1,\ell_2,\ell_3, L_2 ,n_2, n_3}(\Omega_p)  =
   \left [ \Bigl ( Y_{\ell_1}(\hat x_{1p})
    Y_{\ell_2}(\hat x_{2p}) \Bigr )_{L_2}  Y_{\ell_3}(\hat x_{3p}) \right
    ]_{\Lambda M} {\cal P}(\phi_{2p},\phi_{3p}) \ ,\label{eq:hh4}
\end{equation}
where
\begin{eqnarray}
 {\cal P}(\phi_{2p},\phi_{3p}) &= &
 {\cal N}^{\ell_1,\ell_2,\ell_3}_{ n_2, n_3}
   \sin^{\ell_1 }\phi_{2p}    \cos^{\ell_2}\phi_{2p}
   \sin^{\ell_1+\ell_2+2n_2}\phi_{3p}
      \cos^{\ell_3}\phi_{3p} \times    \nonumber \\
   &&
      P^{\ell_1+\frac{1}{2}, \ell_2+\frac{1}{2}}_{n_2}(\cos2\phi_{2p})
      P^{\ell_1+\ell_2+2n_2+2, \ell_3+\frac{1}{2}}_{n_3}(\cos2\phi_{3p})\ ,
      \label{eq:hh4P}
\end{eqnarray}
where $P^{a,b}_n$ are Jacobi polynomials and the coefficients ${\cal N}^{\ell_1,\ell_2,\ell_3}_{ n_2, n_3}$ are normalization factors. The quantity $K=\ell_1+\ell_2+\ell_3+2(n_2+n_3) $ is the so-called grand angular quantum number.  The HH functions are the eigenfunctions of the hyperangular part of the kinetic energy operator. Another important property of the HH functions is that $\rho^K   {\cal  Y}^{K,\Lambda,M}_{\ell_1,\ell_2,\ell_3, L_2 ,n_2, n_3}(\Omega_p)$ are homogeneous polynomials of the particle coordinates of degree $K$.

A set of antisymmetrical hyperangular-spin-isospin states of grand angular quantum number $K$, total orbital angular momentum $\Lambda$, total spin $\Sigma$, total isospin $T$, total angular momentum $J$, and parity $\pi$ can be constructed as follows:
\begin{equation}
  \Psi_{\mu}^{K\Lambda\Sigma TJ\pi} = \sum_{p=1}^{12}
  \Phi_\mu^{K\Lambda\Sigma TJ\pi}(i,j,k,m)\ ,
  \label{eq:PSI}
\end{equation}
where the sum is over the $12$ even permutations $p\equiv i,j,k,m$, and
\begin{equation}
  \Phi^{K\Lambda\Sigma TJ\pi}_{\mu}(i,j;k;m) =  \biggl \{
   {\cal Y}^{K,\Lambda,M}_{\ell_1,\ell_2,\ell_3, L_2 ,n_2, n_3}(\Omega_p)
      \biggl [\Bigl[\bigl[ s_i s_j \bigr]_{S_a}
      s_k\Bigr]_{S_b} s_m  \biggr]_{\Sigma} \biggr \}_{JJ_z}
      \biggl [\Bigl[\bigl[ t_i t_j \bigr]_{T_a}
      t_k\Bigr]_{T_b} t_m  \biggr]_{TT_z}\ .
     \label{eq:PHI}
\end{equation}
Here, ${\cal Y}^{K,\Lambda,M}_{\ell_1,\ell_2,\ell_3, L_2 ,n_2, n_3}(\Omega_p)$ is the HH state defined in~(\ref{eq:hh4}), and $s_i$ ($t_i$) denotes the spin (isospin) function of particle $i$. The total orbital angular  momentum $\Lambda$ of the HH function is coupled to the total spin $\Sigma$ to give a total angular momentum $J$ and parity $\pi=(-1)^{\ell_1+\ell_2+\ell_3}$. The integer index $\mu$ labels the possible choices of hyperangular, spin and isospin quantum numbers, namely
\begin{equation}
   \mu \equiv \{ \ell_1,\ell_2,\ell_3, L_2 ,n_2, n_3, S_a,S_b, T_a,T_b \}\ ,
   \label{eq:mu}
\end{equation}
that are compatible with the given values of $K$, $\Lambda$, $\Sigma$, $T$, $J$ and $\pi$. Another important classification of the states is to group them in ``channels'': states belonging to the same channel have the same values of angular ($\ell_1,\ell_2,\ell_3, L_2 ,\Lambda$), spin ($S_a,S_b,\Sigma$), and isospin ($T_a,T_b,T$) quantum numbers but different values of $n_2$, $n_3$.

Each state  $\Psi^{K\Lambda\Sigma T J\pi}_\mu$ entering the expansion of the $4N$ wavefunction must be antisymmetric under the exchange of any pair of particles. Consequently, it is necessary to consider states such that
\begin{equation}
    \Phi^{K\Lambda\Sigma TJ\pi}_\mu(i,j;k;m)= 
    -\Phi^{K\Lambda\Sigma TJ\pi}_\mu(j,i;k;m)\ ,
     \label{eq:exij}
\end{equation}
which is true when the condition
\begin{equation} 
    \ell_3+S_a+T_a = {\rm odd}\ , \label{eq:lsa}
\end{equation}
is satisfied.

The number $M_{K\Lambda\Sigma TJ\pi}$ of antisymmetrical functions $\Psi^{K\Lambda\Sigma TJ\pi}_\mu$ having given values of $K$, $\Lambda$, $\Sigma$, $T$, $J$ and $\pi$ but different combination of quantum numbers $\mu$ (see Eq.~(\ref{eq:mu})) is in general very large.  In addition to the degeneracy of the HH basis, the four spins (isospins) can be coupled in different ways to $S$ ($T$). However, many of the states $\Psi^{K\Lambda\Sigma TJ\pi}_\mu$, $\mu=1,\ldots,M_{K\Lambda\Sigma TJ\pi}$ are linearly dependent. In the expansion of a $4N$ wave function it is necessary to include the subset of linearly independent states only, whose number is fortunately significantly smaller than the corresponding value of $M_{K\Lambda\Sigma TJ\pi}$.

The internal part of the  wave function can be finally written as
\begin{equation}\label{eq:PSI3}
  \Psi^{LSJJ_z\pi}_C= \sum_{K\Lambda\Sigma T}\sum_{\mu} 
    u^{LSJJ_z\pi}_{K\Lambda\Sigma T,\mu}(\rho)
    \Psi_{\mu}^{K\Lambda\Sigma TJ\pi}\ ,
\end{equation}
where the sum is restricted only to the linearly independent states. 

The main problem is the computation of the matrix elements of the Hamiltonian. This task is considerably simplified by using the following transformation
\begin{equation}\label{eq:arare}
  \Phi^{K\Lambda\Sigma TJ\pi}_{\mu}(i,j;k;m) =
  \sum_{\mu'}  a^{K\Lambda\Sigma TJ\pi}_{\mu,\mu'}(p) 
   \Phi^{K\Lambda\Sigma TJ\pi}_{\mu'}(1,2;3;4)\ .
\end{equation}
The coefficients $a^{K\Lambda\Sigma TJ\pi}_{\mu,\mu'}(p)$ have been obtained using the techniques described in Ref.~\onlinecite{V98}. Then the kinetic energy operator matrix elements are readily obtained analytically, and the \NN ($3N$) potential matrix elements can be obtained by one (three) dimensional integrals.  The details are given in Ref.~\onlinecite{viv05}.

The $S$-matrix elements ${\cal S}^J_{LS,L^\prime S^\prime}(p)$ and functions $u_{\mu}(\rho)$ occurring in the expansion of $\Psi^{LSJJ_z\pi}_C$ are determined by making the functional
\begin{equation}
   [\overline{\cal S}^J_{LS,L^\prime S^\prime}(q)]=
    {\cal S}^J_{LS,L^\prime S^\prime}(q)
     -\frac{M_{N}}{\sqrt{6}\,{\rm i}}
        \left \langle\Psi^{L^\prime S^\prime JJ_z }_{1+3} \left |
         H-E_3-\frac{q^2}{2 \mu} \right |
        \Psi^{LSJJ_z}_{1+3}\right \rangle
\label{eq:kohn}
\end{equation}
stationary with respect to variations in the ${\cal S}^J_{LS,L^\prime S^\prime}$ and $u_{\mu}(\rho)$ (Kohn variational principle).  Here $E_3$ is the $\het$ ground-state energy. By applying this principle, a set of second order differential equations for the functions $u^{LSJJ_z\pi}_{K\Lambda\Sigma T,\mu}(\rho)$ is obtained. By replacing the derivatives with finite differences, a linear system is obtained which can be solved using the Lanczos algorithm.  This procedure, which allows for the solution of a large number of equations, is very similar to that outlined in the Appendix of Ref. \onlinecite{KMRV96} and it will not be repeated here.

The main difficulties of the application of the HH technique are the slow convergence of the basis with respect to the grand angular quantum number $K$, and the (still) large number of linearly independent HH states with a given $K$. Also a brute-force application of the method is not possible even with the most powerful computers available, so one has to select a suitable subset of states~\cite{E72,D77,F83}. In the present work, the HH states are first divided into \emph{classes} depending on the value of $\mathcal{L} = \ell_1 + \ell_2 + \ell_3$, total spin $\Sigma$, and $n_2$, $n_3$.  In practice, HH states of low values of $\ell_1$, $\ell_2$, $\ell_3$ are first included. Between them, those correlating only a particle pair are included first (i.e. those with $n_2=0$), then those correlating three particles are added and so on. The calculation begins by including in the expansion of the wave function the HH states of the first class C1 having grand angular quantum number $K\le K_1$ and studying the convergence of a quantity of interest (for example, the phase-shifts) by increasing the value of $K_1$. Once a satisfactory value of $K_1 = K_{1\rm max}$ is reached, the states of the second class with $K\le K_2$ are added in the expansion, keeping all the states of the class C1 with $K_1 = K_{1\rm max}$. Then $K_2$ is increased until the desired convergence is achieved and so on.

Let us consider, for example, the case $J^\pi=0^-$, where there is only one $LS$ channel in the sum over $L'S'$ of Eq.~(\ref{eq:psia}), namely $L'=1$, $S'=1$.  Consequently, for this wave the S-matrix reduces to a value which is parametrized (as usual) as ${\cal S}^0_{LS,LS} = \eta \exp(2{\rm i} \delta)$, where $\eta$ is known as the ``elasticity parameter''.  Note that in the application of the Kohn principle given in~(\ref{eq:kohn}) the value $\eta = 1$ is not guaranteed: it is achieved only when the corresponding internal part $\Psi_C^{LSJJ_z\pi}$ is well described by the HH basis.  We have used the value of $\eta$ as a check of the convergence of the HH expansion.  In cases of poor convergence, the value of $\eta$ has been found to depend very much also on the choice of $f_L(y_i)$, the function used to regularize the Coulomb function $G_L$.  This function has been chosen to depend on a non-linear parameter $\gamma$, and thus a test of the convergence is performed by analyzing the dependence of $\eta$ on the parameter $\gamma$.  At the beginning of the calculation, when the number of HH functions is not large enough to get convergence, $\eta$ is extremely dependent on the value of $\gamma$.  The phase shift $\delta$, however, depends less critically on $\gamma$.  By increasing the number of HH components in the internal wave function we observe that $\eta \rightarrow 1$, and that the dependence on $\gamma$ becomes negligible.  Notice that the convergence rate has been found to depend on the value of $\gamma$; some critical values of this parameter exist where the convergence can be very slow.  However, it is not difficult to find ranges of values of $\gamma$ where the convergence is fast and smooth and the final results are independent of $\gamma$.  Since we are interested in the convergence of the HH function, we have chosen $\gamma$ in one of the ``favorable'' regions, where the convergence is achieved in a smooth and fast way.  A detailed study on this subject will be reported elsewhere.

Let us now briefly discuss the choice of the classes of HH states for the $J^\pi = 0^-$ case.  Note that, since the wave under consideration is of negative parity, only HH functions with odd values of $K$ and $\mathcal{L}$ have to be considered. Moreover, we consider in this work only states with total isospin $T = 1$, as the effect of the states with $T = 2$ should be negligible.  The criteria used to select the appropriate classes of HH functions require that the states with lowest ${\cal L}$ be considered first.  A few of the channels considered in the calculation have been reported in Table~\ref{table:theory:chan0m1}.  The final choice of classes for the case $J^\pi = 0^-$ is detailed below.

\begin{table}
\begin{ruledtabular}
	\caption[Table]{ Quantum numbers of the first seventeen channels considered in the expansion of the $0^-$ state wave functions. See the text for details.}

	\begin{tabular}{r@{$\quad$}@{$\quad$}r@{$\quad$}
    r@{$\quad$}r@{$\quad$}r@{$\quad$}r@{$\quad$}
    r@{$\quad$}r@{$\quad$}r@{$\quad$}r@{$\quad$}r@{$\quad$}r}
        $\alpha$ & $\ell_1$ & $\ell_2$ & $\ell_3$ & $L_2$ & $\Lambda$ &
        $S_a$ & $S_b$ & $\Sigma$ & $T_a$ & $T_b$ & $T$ \\ \hline
         1 & 1 & 0 & 0 & 1 & 1 & 1 & 1/2 & 1 & 0 & 1/2 & 1\\
         2 & 1 & 0 & 0 & 1 & 1 & 1 & 3/2 & 1 & 0 & 1/2 & 1\\
         3 & 1 & 0 & 0 & 1 & 1 & 0 & 1/2 & 1 & 1 & 1/2 & 1\\
         4 & 1 & 0 & 0 & 1 & 1 & 0 & 1/2 & 1 & 1 & 3/2 & 1\\
         5 & 1 & 0 & 2 & 1 & 1 & 1 & 1/2 & 1 & 0 & 1/2 & 1\\
         6 & 1 & 0 & 2 & 1 & 1 & 1 & 3/2 & 1 & 0 & 1/2 & 1\\
         7 & 1 & 0 & 2 & 1 & 1 & 0 & 1/2 & 1 & 1 & 1/2 & 1\\
         8 & 1 & 0 & 2 & 1 & 1 & 0 & 1/2 & 1 & 1 & 3/2 & 1\\
         9 & 0 & 1 & 0 & 1 & 1 & 1 & 1/2 & 1 & 0 & 1/2 & 1\\
        10 & 0 & 1 & 0 & 1 & 1 & 1 & 3/2 & 1 & 0 & 1/2 & 1\\
        11 & 0 & 1 & 0 & 1 & 1 & 0 & 1/2 & 1 & 1 & 1/2 & 1\\
        12 & 0 & 1 & 0 & 1 & 1 & 0 & 1/2 & 1 & 1 & 3/2 & 1\\
        13 & 0 & 0 & 1 & 0 & 1 & 1 & 1/2 & 1 & 1 & 1/2 & 1\\
        14 & 0 & 0 & 1 & 0 & 1 & 1 & 1/2 & 1 & 1 & 3/2 & 1\\
        15 & 0 & 0 & 1 & 0 & 1 & 1 & 3/2 & 1 & 1 & 1/2 & 1\\
        16 & 0 & 0 & 1 & 0 & 1 & 1 & 3/2 & 1 & 1 & 3/2 & 1\\
        17 & 0 & 0 & 1 & 0 & 1 & 0 & 1/2 & 1 & 0 & 1/2 & 1\\
    \end{tabular}
    \label{table:theory:chan0m1}
\end{ruledtabular}
\end{table}

\begin{enumerate}

	\item {Class C1}. In this class are included the HH states with $n_2 = 0$ belonging to the channels 1 through 8 of Table~\ref{table:theory:chan0m1}. Note that the corresponding radial part of the HH functions depends essentially on $\cos\phi_{3p}=r_{ij}/\rho$ and  thus these states take into account two-body correlations (see Equation~(\ref{eq:hh4P})).  This part of the wavefunction is more difficult to construct due to the strong repulsion at short interparticle distances.

	\item {Class C2}. This class includes HH functions belonging to the same eight channels as for class C1, but with $n_2 > 0$. Since $\cos\phi_{2p}$ is proportional to the distance of particle $k$ from the center of mass of the pair $ij$, these states, therefore, include also part of the three-body correlations.

	\item {Class C3}. This class includes the remaining $T = 1$ states of the channels having $\mathcal{L} = 1$ (channels 9 through 17 of Table~\ref{table:theory:chan0m1}).

	\item {Class C4}. This class includes the $T = 1$  states belonging to the remaining channels with $\mathcal{L} = 3$ and $\Sigma = 1$.

	\item {Class C5}. This class includes the $T = 1$  states belonging to the channels with $\mathcal{L} = 3$ and $\Sigma = 2$.

	\item {Class C6}. This class includes the $T = 1$  states belonging to the channels with $\mathcal{L} = 3$ and $\Sigma = 0$.

	\item {Class C7}. This class includes the $T = 1$  states belonging to the channels with $\mathcal{L} = 5$.

\end{enumerate}

All states of the first four classes have a total spin $\Sigma = 1$. The classification related to the total spin is important since we have observed that the component with $\Sigma = 1$ is the dominant one and requires more states to be well accounted for, while the $\Sigma = 2$ and $\Sigma = 0$ components give only a tiny contribution to the phase shift (however, they are important for achieving $\eta = 1$).

We have also calculated the phase-shifts of the states $J^\pi = 0^+$, $1^+$, $2^+$, $3^+$, $1^-$ and $2^-$. The choice of the classes in these cases has been performed in the same way as discussed above.  In order to test the accuracy reached by the theory, significant work has been done checking the convergence of the HH expansion in terms of the various classes. Some examples of the convergence for the phase shift $\delta$ and elasticity parameter $\eta$ using the AV18 potential are discussed in the Appendix.


\section{Experiments}
\label{sec:exp}

Angular distributions of cross sections \dsig and proton analyzing powers \ay were measured with high precision for proton$-\nuc{3}{He}$ elastic scattering at several energies below 5 MeV.  Silicon surface-barrier detectors, having an effective efficiency at these energies of $100.0 \pm 0.1$\%, were used for both sets of measurements.   In all of the measurements, a pulse generator signal was sent to the test input of the preamplifier of each detector and used as a measure of the dead-time of the data acquisition system.

\begin{table} \centering
\begin{ruledtabular}    
	\caption{Summary of Measurements.  The \ay measurements were made using a proton beam and the \dsig measurements were made using a \nuc{3}{He} beam.  The center-of-mass energies at which the two sets of measurements were made do not exactly agree due to differences in analyzing magnet calibrations for the two sets of experiments, and due to energy losses in the targets.  All energies are in MeV.}

	\begin{tabular}{ddddd} 
		\multicolumn{2}{c}{\ay measurements} & \multicolumn{3}{c}{\dsig measurements} \\
		\multicolumn{1}{c}{\Ecm} & \multicolumn{1}{c}{\Ep} &
		\multicolumn{1}{c}{\Ecm} & \multicolumn{1}{c}{\Ep} & \multicolumn{1}{c}{\Ehe}\\  \hline
		      &          & 0.74  &  0.99  &   2.97 \\
		1.20  &  1.60    & 1.19  &  1.59  &   4.76 \\
		1.69  &  2.25    & 1.67  &  2.24  &   6.69 \\
		2.35  &  3.13    & 2.33  &  3.11  &   9.31 \\
		3.03  &  4.05    & 3.02  &  4.02  &  12.06 \\
	\end{tabular}

	\label{table:exp:energies}
\end{ruledtabular}
\end{table}

\subsection{Cross-section measurements}
\label{sec:exp:dsig}

All of the measurements of the differential cross section were made using the TUNL supersonic gas-jet target, a refurbished and upgraded version of the target designed and built at the University of Erlangen-N\"{u}rnberg~\cite{bit79}.  Target thicknesses of $\approx 5 \times 10^{17}$ atoms/cm$^{2}$ were routinely obtained in a narrow jet ($\approx$ 1 mm) for the measurements in this work, facilitating very reasonable counting times for the cross-sections measured.  The target's lack of beam-degrading windows, well-defined geometry, and high-purity and high-density were ideal for these low-energy measurements. 

\begin{figure}
	\includegraphics[width=\columnwidth]{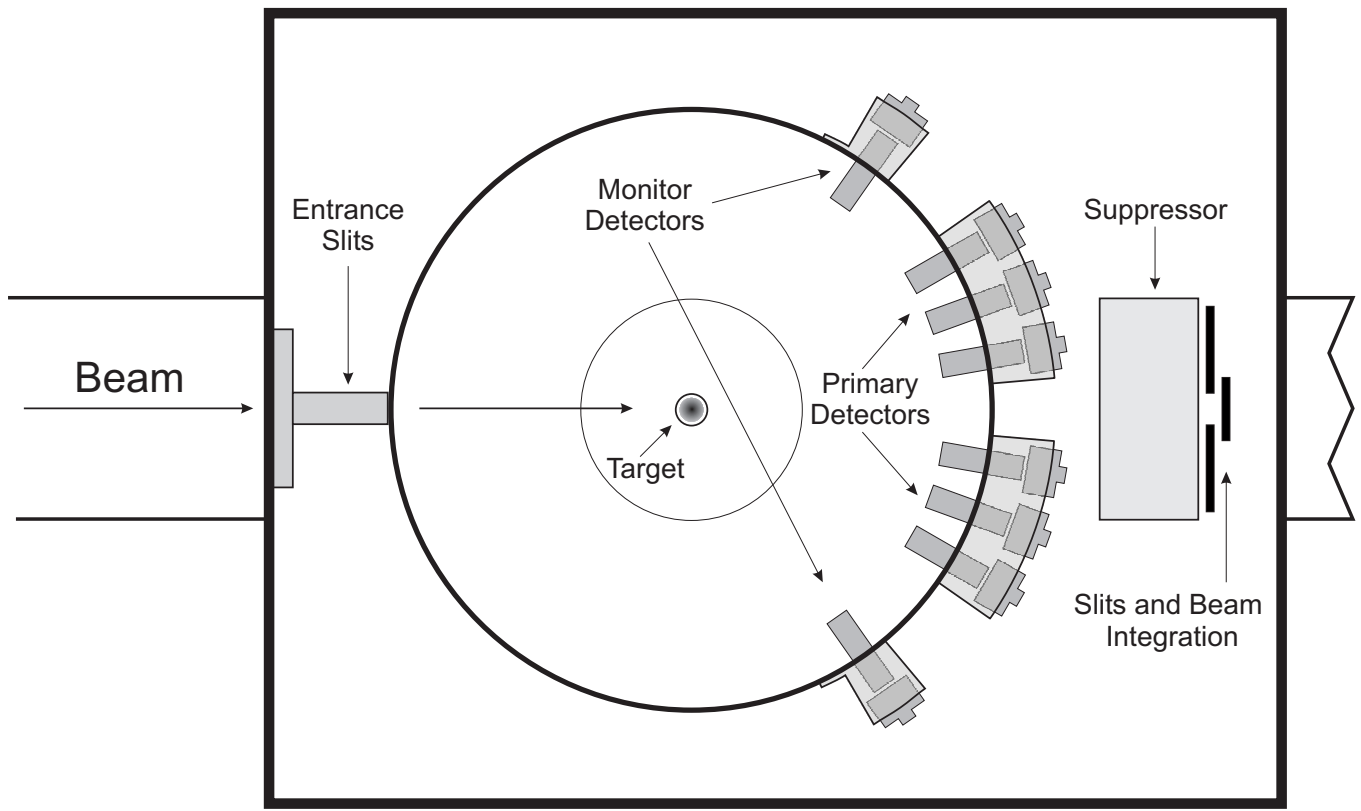}

	\caption{Scattering chamber setup for \dsig measurements.  The primary detectors are arranged symmetrically about the beam direction.  For the relative \dsig measurements, the yields in the primary detectors were normalized to the counts in the fixed monitor detectors shown.  The beam position and beam current were monitored using a set of slits behind the target.}

	\label{fig:dsig:chambersetup}
\end{figure}

\begin{figure}
	\includegraphics[width=\columnwidth]{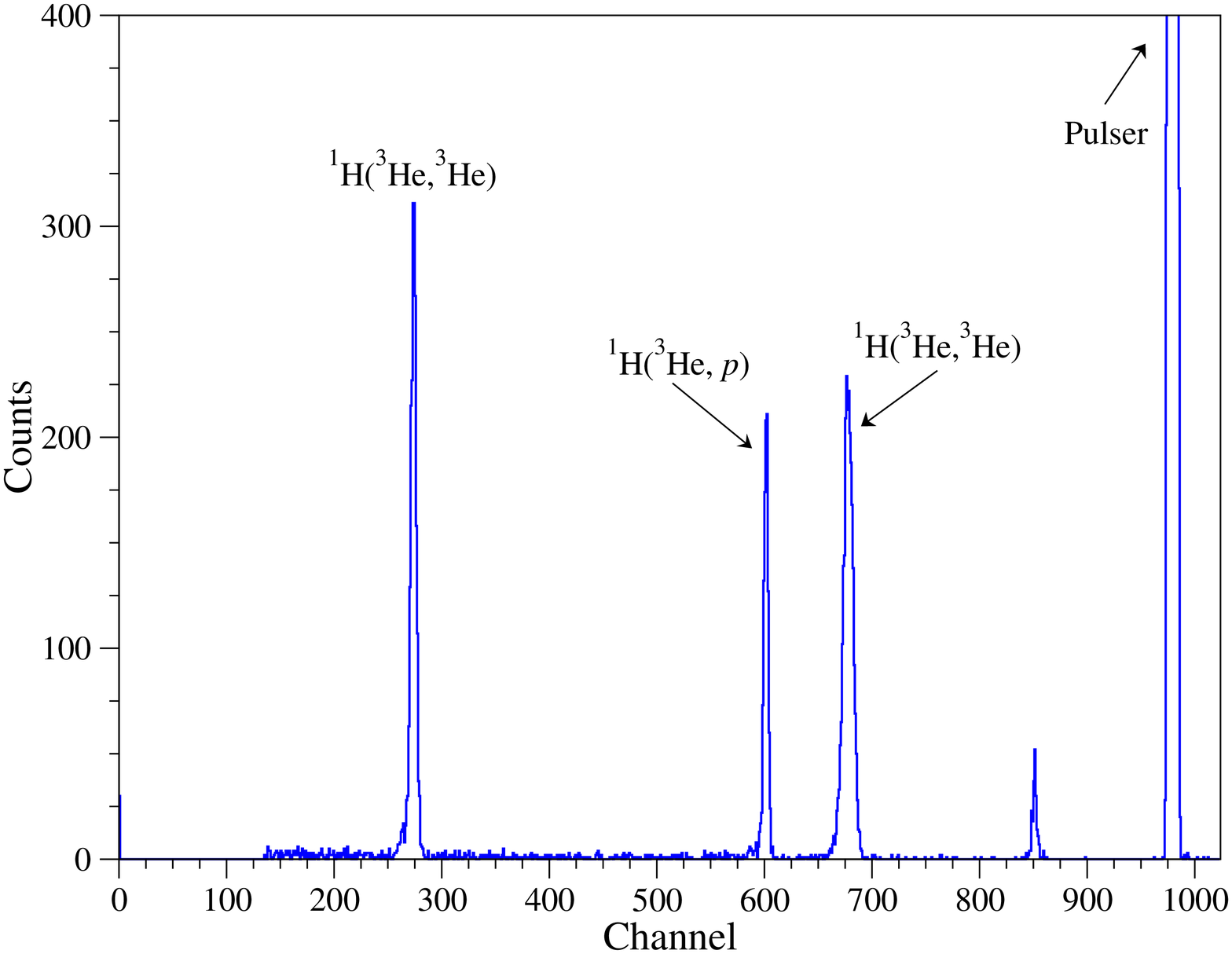}

	\caption{(Color online) Typical energy spectrum of particles from the scattering of the \nuc{3}{He} beam on the hydrogen target.  This spectrum was measured at $\tlab = 15\degree$ at a \nuc{3}{He} beam energy of 6.73 MeV.  The peak at channel 850 is caused by heavier gas contaminants.}

	\label{fig:dsig:rel-spec}
\end{figure}

Since \nuc{3}{He} is a very expensive gas, the measurements were performed in inverse kinematics using a \nuc{3}{He} beam incident on a hydrogen gas-jet.  Measurements of \dsig were made at the five \nuc{3}{He} energies listed in Table \ref{table:exp:energies}.  Scattered \nuc{3}{He} particles can only be detected forward of $\tlab \approx 19\degree$.  Detection of both the scattered \nuc{3}{He} and the recoiling proton allowed for the measurement of \dsig over a wide range of center of-mass-angles.

\begin{table} \centering
\begin{ruledtabular}
	\caption{Collimation configurations for the relative \dsig measurements.  All lengths are in mm.  All detectors were 16.6 cm from the target center, and fitted with a 5.1 cm snout.  H is the horizontal collimator dimension and V is the vertical collimator dimension.  The ``Forward Angles'' column shows the detector setups for forward angle measurements (with the monitors at 55\degree) and the ``Backward Angles'' column shows the setup for backward angle measurements (with the monitors at 15\degree.)}

\begin{tabular}{cccccc}
           &          & \multicolumn{2}{c}{\underline{Forward Angles}} & \multicolumn{2}{c}{\underline{Backward Angles}} \\
    \Ehe   & Detector & Front        &  Rear         & Front        &  Rear         \\
    (MeV)  &          & H $\times$ V  (mm$^{2}$) &  H $\times$ V (mm$^{2}$) & H $\times$ V (mm$^{2}$) &  H $\times$ V (mm$^{2}$) \\ \hline
    2.97   & Primary  & 2.4 \by 9.5  &  0.8 \by 9.5  & 2.4 \by 9.5  &  0.8 \by 9.5  \\
           & Monitor  & 6.4 \by 9.5  &  6.4 \by 9.5  & 2.4 \by 9.5  &  0.8 \by 9.5  \\ \hline
    4.76   & Primary  & 2.4 \by 9.5  &  0.8 \by 9.5  & 3.2 \by 9.5  &  1.6 \by 9.5  \\
           & Monitor  & 6.4 \by 9.5  &  6.4 \by 9.5  & 2.4 \by 9.5  &  0.8 \by 9.5  \\ \hline
    6.69   & Primary  & 2.4 \by 9.5  &  0.8 \by 9.5  & 2.4 \by 9.5  &  0.8 \by 9.5  \\
           & Monitor  & 6.4 \by 9.5  &  6.4 \by 9.5  & 2.4 \by 9.5  &  0.8 \by 9.5  \\ \hline
    9.31   & Primary  & 2.4 \by 9.5  &  0.8 \by 9.5  & 3.2 \by 9.5  &  1.6 \by 9.5  \\
           & Monitor  & 6.4 \by 9.5  &  3.2 \by 9.5  & 2.4 \by 9.5  &  0.8 \by 9.5  \\ \hline
    12.06  & Primary  & 2.4 \by 9.5  &  0.8 \by 9.5  & 2.4 \by 9.5  &  0.8 \by 9.5  \\
           & Monitor  & 6.4 \by 9.5  &  6.4 \by 9.5  & 2.4 \by 9.5  &  0.8 \by 9.5  \\
\end{tabular}
\label{table:dsig:detectors}
\end{ruledtabular}
\end{table}

For each energy, a beam of \nuc{3}{He}$^{++}$ ions was accelerated with the FN tandem accelerator and then deflected onto the hydrogen gas-jet target. The elastically scattered protons and \nuc{3}{He} nuclei were detected by three pairs of primary detectors placed symmetrically about the beam direction, as shown in Fig. \ref{fig:dsig:chambersetup}.  Each detector was fitted with a pair of rectangular slits mounted in a cylindrical ``snout.''  The dimensions of these slits are given in Table \ref{table:dsig:detectors}.   The counts in each detector were normalized to the yield of scattered particles in a pair of monitor detectors also placed symmetrically about the scattering region.  The angular range covered with the movable chamber detectors was $7.5\degree$ to $75\degree$.  When measuring forward angles the monitor detectors were placed at $55\degree$, while for more backward angle measurements the monitor detectors were placed at $15\degree$.  A cross-normalization was performed to maintain a consistent normalization for both sets of monitor detector angle settings.  A typical spectrum is shown in Fig. \ref{fig:dsig:rel-spec}.  The beam position relative to the jet-target was monitored by a pair of horizontal and vertical slits behind the target region. The beam passing through the small slit opening was integrated by a Faraday cup which was electrically isolated from the slits.  For each measurement the beam current was maximized on the Faraday cup, thus making sure that the scattering geometry remained the same from run to run.  The currents from the slits and Faraday cup were summed and used to measure the total integrated charge on target.

At lower beam energies, multiple scattering in the gas-jet decreased the measured total integrated charge and a small correction factor had to be applied.  This factor was determined by frequently cycling the gas in the jet on and off and determining the effect of the gas presence on the integrated beam current.  This was found to be a $(6 \pm 1)$\% correction at $\Ehe = 2.97$ MeV, reducing to a negligible correction at $\Ehe = 12$ MeV.

The absolute normalizations of the \dsig measurements were performed with two methods. In the first method, proton-\nuc{3}{He} elastic scattering was normalized to \nuc{3}{He}-\nuc{40}{Ar} Rutherford scattering.  Bombarding a gas-jet containing both hydrogen and argon with a \nuc{3}{He} beam, the ratio of the proton-\nuc{3}{He} yield to the \nuc{3}{He}-\nuc{40}{Ar} yield in the same detector at a given angle was measured.  If the ratio $R_{t}$ of hydrogen target thickness to argon target thickness is known, the ratio of \phe counts to \nuc{3}{He}-\nuc{40}{Ar} counts at the same angle gives an absolute determination of \dsig at that angle.

For measurements of the absolute normalization using this method, a small amount ($\sim$~3\%) of Argon was mixed with the hydrogen gas making the target jet.  The gas was mixed by the manufacturer and the Ar/H$_{2}$ ratio  $R_{t}$ was determined to an accuracy of 2\% by gas chromatography \cite{NSG}.  $R_{t}$ was also measured using a proton beam at $\Ep = 2.24$ MeV, at angles where proton-\nuc{40}{Ar} elastic scattering is known to be described by the Rutherford formula within a percent, as calculated using several different sets of optical model parameters~\cite{OM}.   Using the well-known proton-proton elastic-scattering cross-section, which was obtained from the high-accuracy phase-shift analysis of the Nijmegen group \cite{ren99, said}, determinations of $R_{t}$ using this method agreed within error with the gas-chromatography measurements.

Similarly, this mixed-jet method also relies on \nuc{3}{He}-\nuc{40}{Ar} elastic scattering being described correctly by the Rutherford scattering formula.  Using the DWBA code \textsc{dwuck4} \cite{dwuck} and optical model parameters from Ref. \onlinecite{bre75}, it was determined that the \nuc{3}{He}-\nuc{40}{Ar} elastic scattering cross-section is within 5\% of the Rutherford prediction out to $\tlab = 40\degree$ for the three lowest \nuc{3}{He} energies in Table \ref{table:exp:energies}.

The other technique for determining the absolute normalization of the \dsig angular distributions was a beam-switching method, in which the product of detector solid-angle and target-thickness was determined using a proton beam incident on a hydrogen jet, via the known proton-proton elastic scattering cross section.  A \nuc{3}{He} beam at the proper energy was first tuned onto the hydrogen jet target; this was followed by irradiating the jet with a proton beam with the same magnetic rigidity.  This procedure allowed for minimal beam-transport adjustments; the source inflection magnet before the tandem accelerator and the accelerator terminal potential were adjusted so that both beams passed into the chamber with the same beam tune.  This ensured the beam-target geometry was the same for each beam. Scattered particles from each beam were detected by three pairs of fixed-angle detectors.  This procedure was repeated several times to ensure reproducibility.

The beam-switching technique was used at energies at which the mixed-jet method was not feasible.  Both techniques were used at several energies, as a cross-check, and the results from both methods agreed within errors.  Typical error budgets for each method are shown in Table \ref{table:exp:errors}, and the overall systematic normalization errors are listed in Table \ref{table:exp:abserrors}.

\begin{table}\centering
\begin{ruledtabular}
	\caption{Typical error budgets for both the mixed-jet and beam-switching absolute normalization methods.  Statistical and systematic errors are listed.}

	\begin{tabular}{lcc}

        \multicolumn{3}{c}{Mixed-Jet} \\
        Source                              & Type  & Error (\%) \\ \hline
        Counting Statistics                 & stat. &  0.3       \\
        \nuc{40}{Ar}/H$_{2}$ Ratio          & stat. &  0.6       \\
                                            & sys.  &  0.5       \\
        \nuc{40}{Ar} - \nuc{3}{He}   \dsig  & sys.  &  0.8       \\
        Angle Setting                       & sys.  &  1.0       \\ \hline
                                            &       &            \\
        \multicolumn{3}{c}{Beam-Switching} \\
        Source                              & Type  & Error (\%) \\ \hline
        Counting Statistics                 & stat. &   0.4      \\
        Proton-proton $\sigma(E,\theta)$    & sys.  &   0.7      \\
        BCI Correction Factor               & sys.  &   1.0      \\
        Beam Energy                         & sys.  & $<$0.1     \\
        Angle Setting                       & sys.  & $<$0.1     \\
        Jet Reproducibility                 & sys.  &   0.8      \\

	\end{tabular}

	\label{table:exp:errors}
\end{ruledtabular}
\end{table}

\begin{table}\centering
\begin{ruledtabular}
	\caption{Overall systematic normalization error for each of the \dsig measurements.}

	\begin{tabular}{cc}
        Equivalent \Ep [MeV]   &  Error (\%) \\ \hline
	0.99                   &  3.5 \\
	1.59                   &  2.0 \\
	2.25                   &  2.7 \\
	3.11                   &  2.9 \\
	4.02                   &  2.7 \\
	\end{tabular}

	\label{table:exp:abserrors}
\end{ruledtabular}
\end{table}

\subsection{Analyzing power measurements}
\label{sec:exp:ay}

\begin{figure} \centering
	\includegraphics[width=\columnwidth]{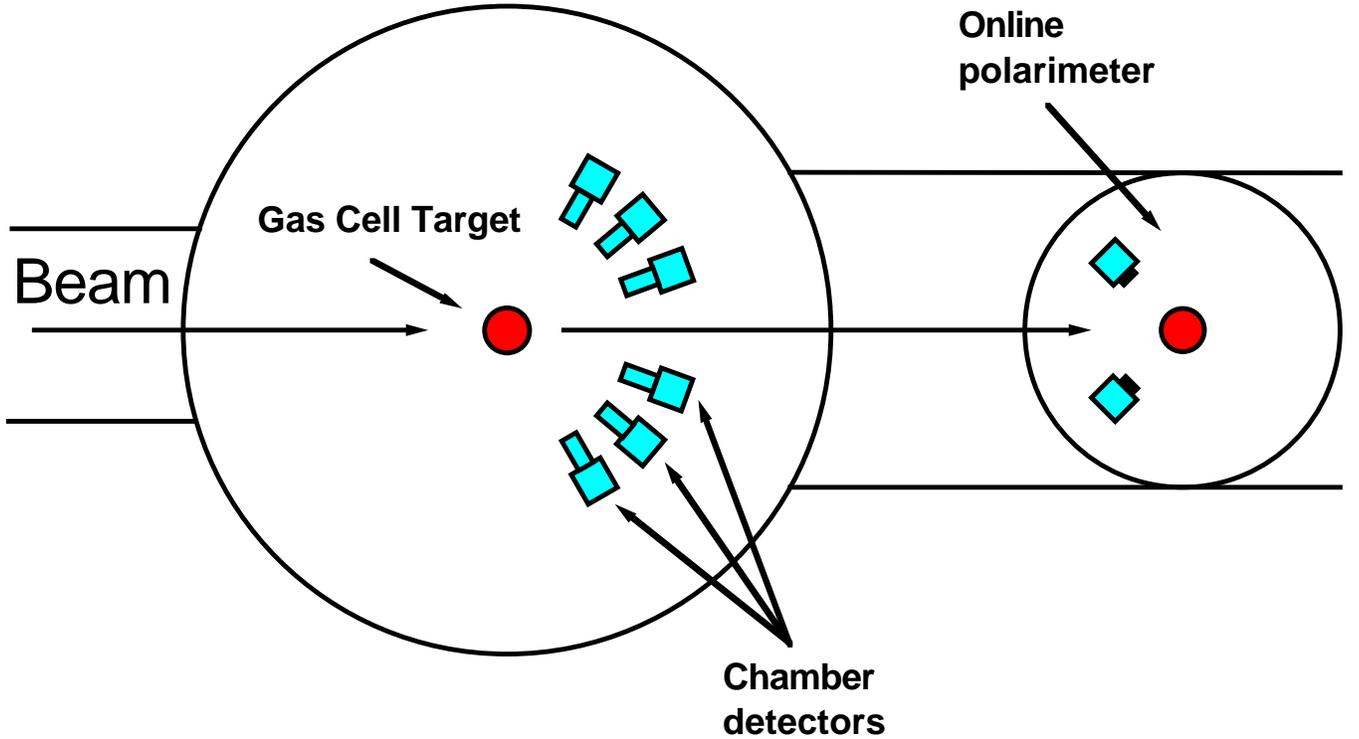}

	\caption{(Color online) Chamber setup for \ay measurements.}

	\label{fig:ay:ay-chamber}
\end{figure}

The measurements of \ay were made utilizing the atomic beam polarized ion source at TUNL \cite{cle95} via a two polarization state method~\cite{bru97} with fast state switching \cite{gei98}.  This polarized proton beam was accelerated to the desired energy with the tandem accelerator and tuned onto a gas-cell target inside the 62 cm diameter scattering chamber.  The gas-cell, employing a 2.3 $\mu$m Havar foil, was filled with 1 atm of \nuc{3}{He} gas, and was mounted on the target-rod which was supported from the top of the chamber.  This allowed the cell to be raised, allowing the beam to directly enter the polarimeter.  A schematic of the experimental setup for the \ay measurements is shown in Fig. \ref{fig:ay:ay-chamber}, and the collimation setup is detailed in Table \ref{table:ay:detectors}.  These collimators limited the view of the detectors to only protons scattered from \nuc{3}{He} gas in the cell and not those scattered from the cell entrance and exit foils.  \ay data was taken only at angles for which foil-scattering was negligible.

\begin{table*}
\begin{ruledtabular} \centering
	\caption{Horizontal (H) and vertical (V) detector slit dimensions for \ay measurements.  The front and back collimators for each detector were the same dimensions, as listed.  The distance from the target center to the detector face $R$, and the length of the detector snout $S$ are also listed.  The detector pairs were spaced at 15\degree intervals.}

\begin{tabular}{cccccccccc} 
      & \multicolumn{3}{c}{Forward Pair} & \multicolumn{3}{c}{Middle Pair} & \multicolumn{3}{c}{Backward Pair} \\ \hline
\Ep   &  H $\times$ V  &  $R$ & $S$  & H $\times$ V  & $R$  & $S$  & H $\times$ V  & $R$  & $S$ \\
(MeV) &   (mm${^2}$)   & (cm) & (cm) &    (mm${^2}$) & (cm) & (cm) & (mm${^2}$) & (cm) & (cm) \\ \hline
    1.60   &  1.6 $\times$ 9.5  &  12   & 6.4   & 1.6 $\times$ 9.5 &  12 & 5.1   & 2.4 $\times$ 9.5 &  14 & 5.1  \\
    2.25   &  1.6 $\times$ 9.5  &  10.2 & 5.1   & 1.6 $\times$ 9.5 &  12 & 5.1   & 2.4 $\times$ 9.5 &  14 & 5.1  \\
    3.13   &  1.6 $\times$ 9.5  &  10.2 & 5.1   & 1.6 $\times$ 9.5 &  12 & 5.1   & 2.4 $\times$ 9.5 &  14 & 5.1  \\
    4.05   &  1.6 $\times$ 9.5  &  10.2 & 5.1   & 1.6 $\times$ 9.5 &  12 & 5.1   & 2.4 $\times$ 9.5 &  14 & 5.1  \\

    \end{tabular}
\label{table:ay:detectors}
\end{ruledtabular}
\end{table*}

Since there is significant energy loss (particularly at the lowest energies) in the cell foil, the incident beam energy was adjusted so that the beam reached the desired energy at the cell center.  The energy losses in the cell foil and \nuc{3}{He} gas were modeled by the computer code \textsc{srim-2000}~\cite{zig}.  The proton energies at the center of the gas-cell are listed in Table \ref{table:exp:energies}.

The polarization of the proton beam was monitored on-line with a polarimeter based on \reaction{4}{He}{$\vec{\mathrm{p}}$}{p}{4}{He} elastic scattering~\cite{sch71}.  Periodically during the experimental runs the beam energy was raised either to 6 MeV or 8 MeV, where the analyzing power for the polarimeter is very close to unity.  This was done once every two to three hours.  The polarization state of the beam was switched approximately three times a second and was typically $\approx (67 \pm 2)$\%.  A 2\% systematic error on the \ay measurements arises from the error in beam polarization determinations.

\section{Comparisons with Theory}
\label{sec:compare}

In this section, the experimental data are presented and compared with the results of the theory reviewed in Sec.~\ref{sec:theory}.  The results for the differential cross sections and analyzing powers are presented in Section~\ref{sec:results:theory-dsig} and Section~\ref{sec:results:theory-ay}, respectively.  Note that the \dsig data are designated by their equivalent proton lab energy \Ep, despite the data being taken in inverse kinematics.  Finally, the theoretical predictions of other observables reported in Ref.~\cite{all93a} are presented and discussed in Section~\ref{sec:results:theory-oo}.  The calculations presented were performed using the Argonne $v_{18}$~\cite{wir95} \NN potential (AV18 model), and with the $v_{18}$ \NN potential with the inclusion of the Urbana IX $3N$ force~\cite{pud97} (AV18/UIX model).  The corresponding phase shift and mixing angle parameters calculated with the HH expansion have reached a noticeable degree of convergence, as discussed in great detail in the Appendix.

\subsection{Differential Cross-Section $\boldsymbol{\dsig}$}
\label{sec:results:theory-dsig}

\begin{figure*}
	\includegraphics[width=\textwidth]{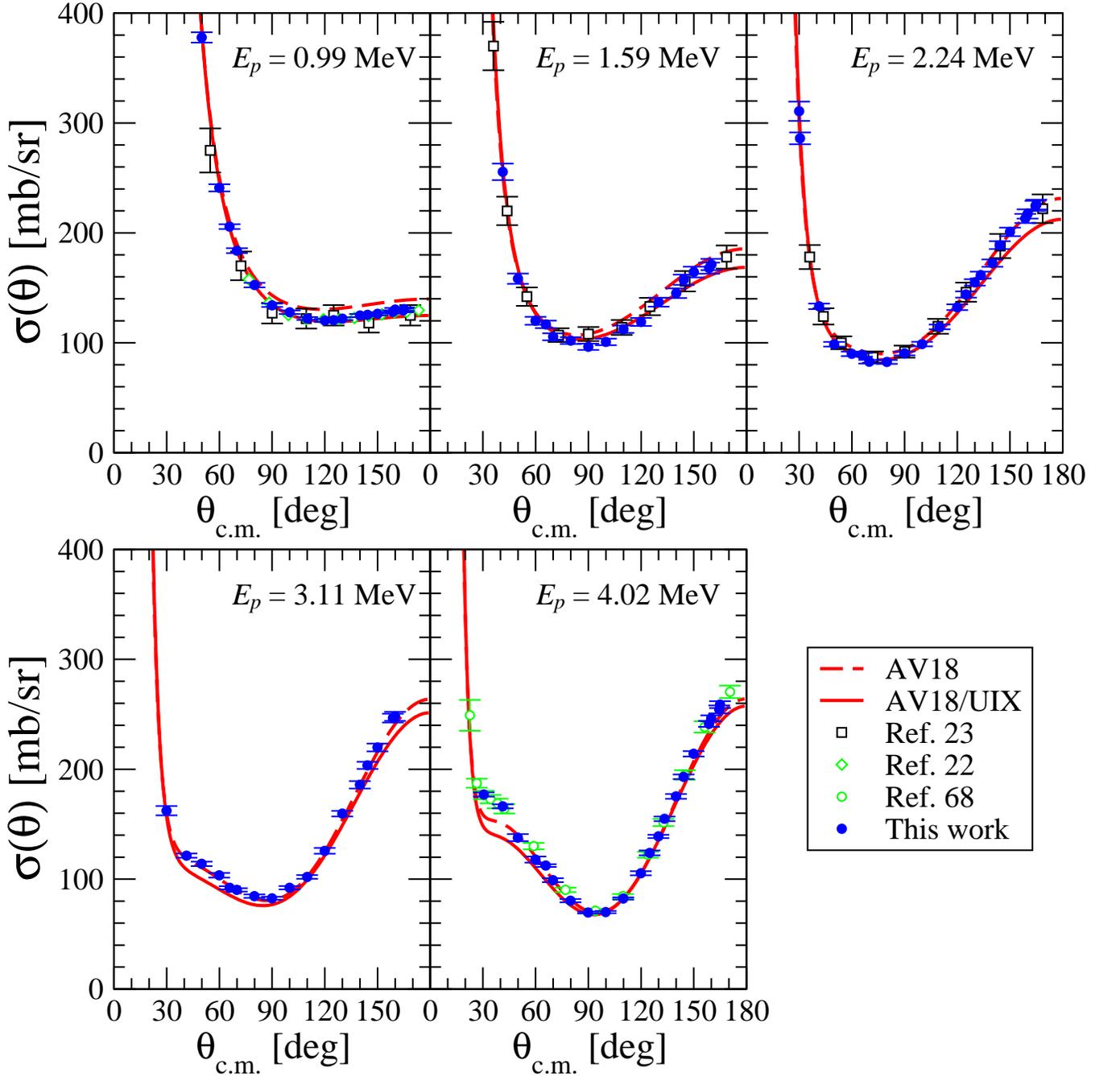}

	\caption{(Color online) The measured \phe elastic differential cross sections (solid circles) at five different energies are compared with the data reported in Ref. \protect{\onlinecite{fam54}} (open squares), Ref. \protect{\onlinecite{ber80}} (open diamonds), and Ref. \protect{\onlinecite{McD64}} (open circles). The curves show the results of the theoretical calculations for the AV18 (dashed lines) and AV18/UIX (solid lines) potential models.}

	\label{fig:dsig}
\end{figure*}

The measured differential cross sections \dsig at the five energies considered here are presented and compared with the existing data~\cite{fam54} in Fig. \ref{fig:dsig}.   The results of the previously described calculations for the AV18 potential (dashed lines) and AV18/UIX model interactions (solid lines) are also shown.

When comparing the data of this work with previous measurements of \dsig at the same energies, the agreement (in general) is quite good; there is a marked improvement in the size of the error bars, and the new data sets contain many more data points.  The precision is much better than the data of Reference \cite{fam54}, slightly better than that of Reference \cite{McD64}, and is comparable to that of Reference \cite{ber80}.  At $\Ep = 4.02$ MeV, there is good agreement between the current data and those of Reference \cite{McD64} but with slightly smaller error bars.  There is no previous data known to exist at $\Ep = 3.11$ MeV.

As can be seen in Fig. \ref{fig:dsig}, there is a general agreement between the theoretical and the experimental results. For small angles, the cross section is dominated by the Coulomb scattering and a good agreement is observed (the exception being two points at small $\tcm$ at $\Ep = 4.02$ MeV).  At $\tcm = 90\degree$, the contribution of the $L = 1$ waves vanishes, and therefore the cross section is almost completely due to the $L = 0$ phase-shifts. As discussed in the Appendix, there are no problems in the calculation of the $L = 0$ phase-shifts from the numerical point of view, and therefore $\sigma(90\degree)$ is an unambiguous test of the underlying nuclear dynamics. We observe a sizeable $3N$ force effect in this region (the minimum), which tends to decrease as \Ep is increased.  This is consistent with the increased binding of the \nuc{3}{He} when the $3N$ force is included.

As $\tcm$ approaches $180\degree$, the predicted cross section becomes quite sensitive to the $L = 1$ phase-shifts (states with  $J^\pi = 0^-$, $1^-$, and $2^-$).  The calculations slightly underestimate the cross sections there, particularly when the $3N$ force is included.  This problem is somewhat analogous to that found in $n$-\nuc{3}{H} elastic scattering in the peak region at $E_n \approx 3$ MeV~\cite{laz05}, as mentioned in Section \ref{sec:intro:nt}.  In that case, the calculations based on the standard \NN and $3N$ forces (as used here) are found to underestimate the total cross section by 20\% on the peak~\cite{laz05}.  These problems probably arise from an incomplete knowledge of either the \NN or the $3N$ interaction in $P$-waves, and therefore are closely related to the \Nd \ay puzzle.  This becomes more evident in the study of the \phe analyzing powers presented below.

\subsection{Proton Analyzing Power $\boldsymbol{\ay}$}
\label{sec:results:theory-ay}

\begin{figure*}
	\includegraphics[width=\textwidth]{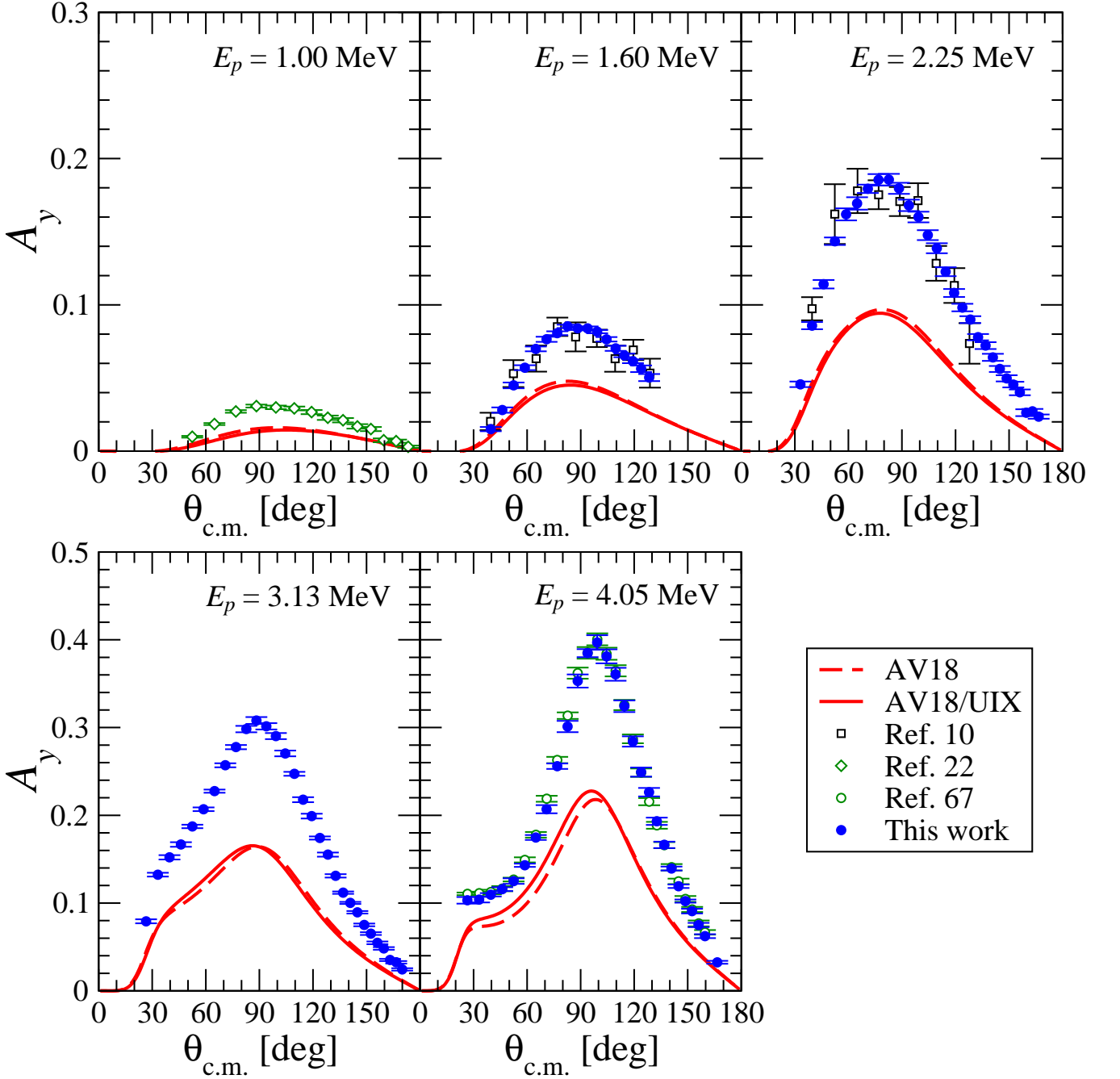}

	\caption{(Color online) The measured \phe proton analyzing power \ay (solid circles) at five different energies are compared with the data of Ref. \protect{\onlinecite{viv01}} (open squares), Ref. \protect{\onlinecite{ber80}} (open diamonds), and Ref. \protect{\onlinecite{all93a}} (open circles). The curves show the results of the theoretical calculations for the AV18 (dashed lines) and AV18/UIX (solid lines) potential models.}

	\label{fig:ay}
\end{figure*}

The measured proton analyzing power \ay at the four energies considered here are presented and compared with the existing data in Fig.~\ref{fig:ay}.  Additionally, at $\Ep = 1.0$ MeV the experimental data of Ref. \onlinecite{ber80} are shown.  The calculations obtained with the AV18 (dashed lines) and AV18/UIX (solid lines) models are also shown.  Note that \ay steadily grows as \Ep is increased.  There is good agreement between the new measurements and the older ones reported in Refs.~\cite{all93a, viv01}.  Note, however, that the present measurements are noticeably more precise, in particular at $\Ep = 1.60$ and $2.25$ MeV.

\begin{figure}
	\includegraphics[width=\columnwidth]{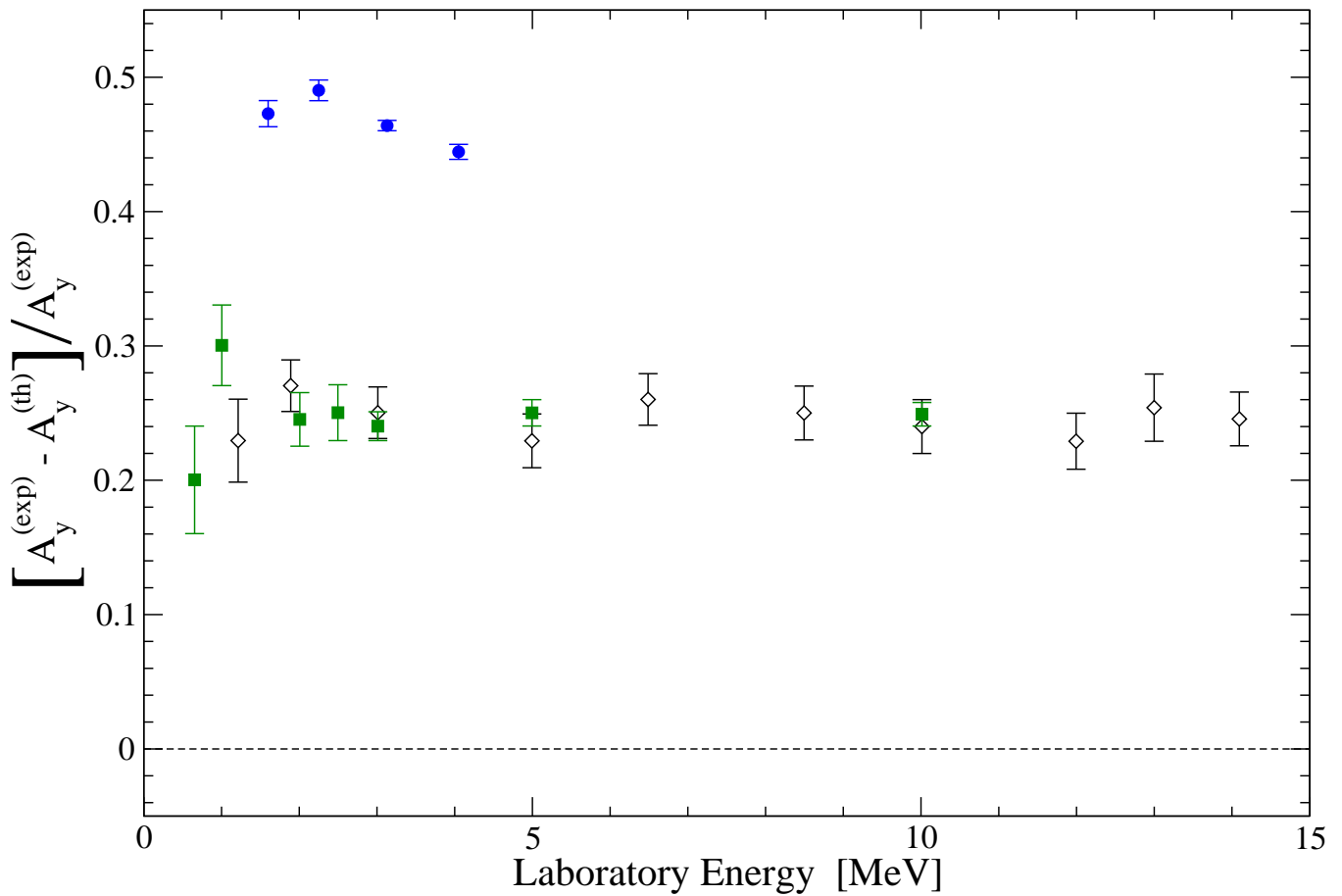}

	\caption{(Color online) Relative difference as a function of energy between the theoretical predictions and measured values for \ay at the peak of the distribution for \phe scattering ($\bullet$).  Also shown are results from Ref. \onlinecite{kie04} for \pd ($\blacksquare$) and \nd scattering ($\diamondsuit$).}

	\label{fig:ay-discrep}
\end{figure}

The theoretical calculations clearly underestimate the data at all energies. The $3N$ force of the Urbana-type has a very little effect at low energies, but its effects are larger at $\Ep = 3.13$ and $4.05$ MeV.  They are, however, clearly insufficient to resolve the discrepancies with the data.  The present results confirm the disagreement previously reported in Refs.~\cite{viv01,pfi01}.  A plot of the relative difference between experiment and theory at the maximum \ay value in the angular distribution as a function of proton energy is shown in Fig.~\ref{fig:ay-discrep}.  Note that this difference is nearly constant as the energy is changed.  This is similar to what is observed in \Nd scattering, though the discrepancy in the \phe case is about 50\% larger.  The \ay observable is very sensitive to the $L = 1$ phase-shifts, and in particular to the combination of phase shifts $\Delta = \delta({}^3P_2) - [\delta({}^3P_1) + \delta({}^3P_0)]/2$~\cite{viv01}.  The value of $\Delta$ is predicted (using the AV18 and AV18/UIX models) to be smaller than the one extracted from the data. It is interesting to note that this is analagous to the \Nd case, in which the splitting between the ${}^4P_{1/2}$ phase and the average of the ${}^4P_{3/2}$ and ${}^4P_{5/2}$ phases is too small to reproduce the observed \ay.  It would be very interesting to see if new terms in the $3N$ interaction could explain both the \Nd and \phe \ay discrepancies.  Work in this direction is in progress.

\subsection{Other Observables at $\Ep = 4.05$ MeV} 
\label{sec:results:theory-oo}

\begin{figure}
	\includegraphics[width=\columnwidth]{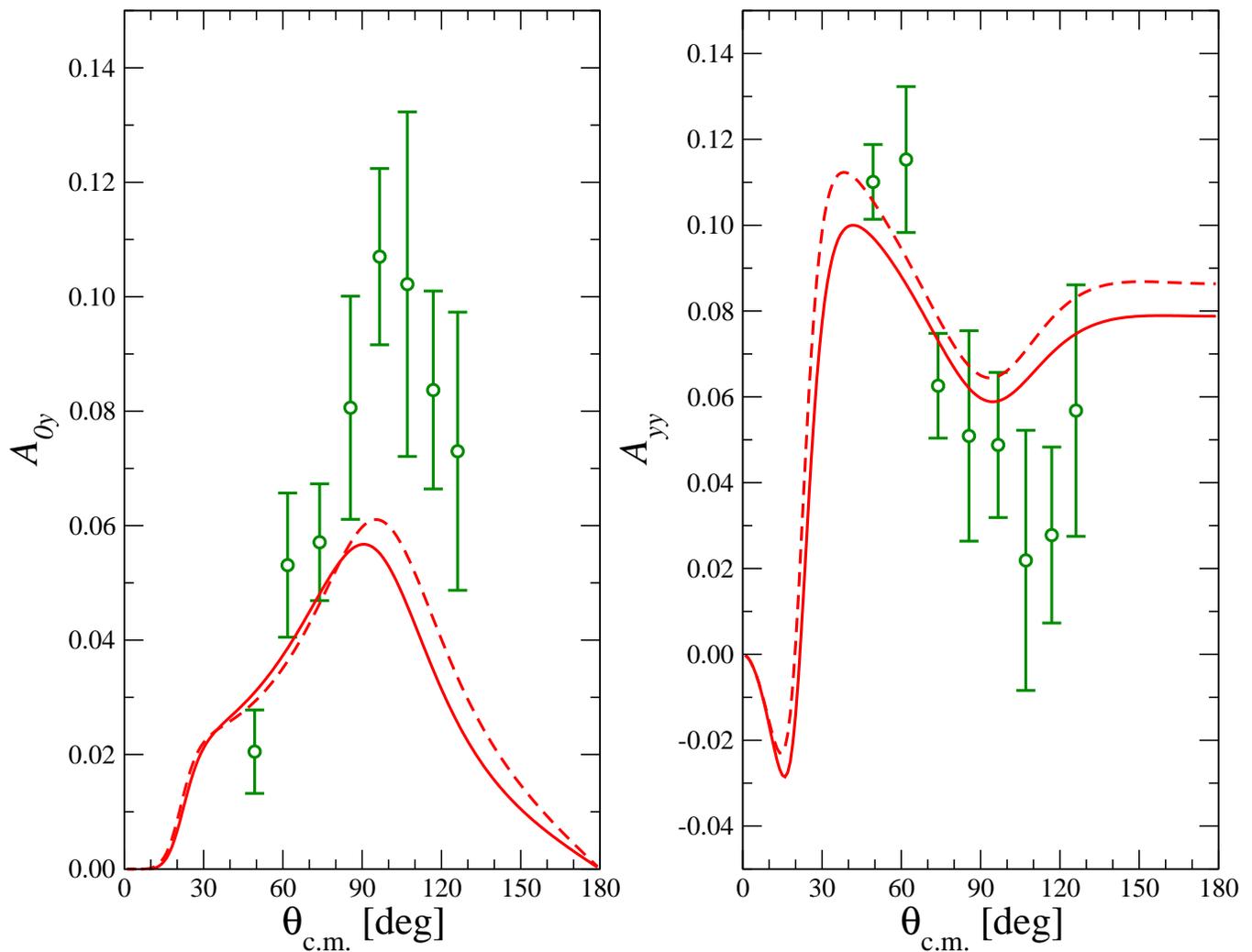}

	\caption{(Color online) \nuc{3}{He} analyzing power \ayhe and spin-correlation coefficient \ayy at $\Ep = 4.05$ MeV.  Data are from Ref. \protect{\onlinecite{all93a}}.  The curves show the theoretical predictions from the AV18 (dashed lines) and AV18/UIX (solid lines) potential models.}

	\label{fig:oo}
\end{figure}

At $\Ep = 4.05$ MeV, measurements of other \phe observables (the spin correlation coefficient \ayy and the \nuc{3}{He} analyzing power \ayhe) exist~\cite{all93a}.  The comparison between the results of the present calculation and these data are shown in Fig.~\ref{fig:oo}.  The measurements have rather large error bars, and no clear conclusion about the agreement with theory for the \ayy measurements can be reached.  However, \ayhe does appear to be under-predicted at the maximum.  Indeed, \ayhe is particularly sensitive to the combination $\delta({}^3P_2) - \delta({}^3P_1)$.  On the other hand, the observable $A_{yy}$ is quite sensitive to $\epsilon(1^+)$, the mixing parameter of the $J^\pi = 1^+$ state.  More precise measurements of these observables could provide much-needed input for an experimental phase-shift analysis and therefore permit a better understanding of the discrepancy with theoretical calculations.  Such measurements are currently underway~\cite{dan05}.

\section{Conclusions}
\label{sec:conclusions}

In this paper the solution of the Schr\"{o}dinger equation for four-nucleon scattering states has been obtained using the HH function expansion.  The main difficulty when using the HH basis is its large degeneracy.  Accordingly, a judicious selection of the HH functions giving the most important contributions has been performed.  For this work, the HH functions have been divided into classes, depending on the number of correlated particles, the values of the orbital angular momenta, the total spin quantum number, etc.  For each class, the expansion has been truncated so as to obtain the required accuracy.  The study of the convergence of \phe elastic scattering phase-shifts and observables reported in the Appendix has shown that good accuracies are achieveable and a powerful method to extrapolate the results has been also discussed.  When applied previously for \nt elastic scattering, the HH method has been proved to give results in good agreement with other theoretical techniques \cite{laz05}.

We also reported measurements of the cross-section \dsig and proton analyzing power \ay for \phe elastic scattering over the range of energies 1.6 MeV $\leq \Ep \leq$ 4.05 MeV.  Additionally, \dsig measurements were obtained at \Ep = 0.99 MeV. Analyzing powers were large and increased in magnitude by more than a factor of 3 as the energy was increased from 1.6 to 4.05 MeV.  Both the \ay and \dsig measurements have higher statistical precision at more angles, and smaller and well-understood systematic errors than those existing previously.

There is good agreement between the cross section data and the calculations when the $3N$ potential is included. However, there are large differences between the \ay data and theory. At the maxima of the \ay angular distribution, the theory underpredicts experimental values by about 50 \%. The inclusion of the $3N$ potential produces only a small percentage change in the predicted analyzing powers and hence has little effect on the magnitude of the disagreement. This disagreement is remarkably similar to (and twice as large as) the ``$A_{y}$ puzzle'' observed for nearly 30 years for \Nd scattering.

The present calculations were extended to include \ayy and \ayhe for which there are measurements at 4.05 MeV \cite{all93a}. The inclusion of the $3N$ force has some influence on predictions of both \ayhe and \ayy.  The calculations for these two observables are much closer to the experimental data although the data have much larger errors than for \ay.  More precise measurements of \ayhe and \ayy could help to define the phase shifts and provide a better understanding of the origin of this new ``$A_{y}$ puzzle.''  Such measurements are currently underway~\cite{dan05}.



\appendix*

\section{Convergence of the calculated phase-shifts}
\label{sec:results:conv}

In this Appendix, we discuss the convergence of the calculated phase-shifts.  At the energies of interest here, \phe scattering is dominated by the $L = 0$ and $1$ waves.  The convergence of the HH expansion of $\Psi_C^{LSJJ_z\pi}$ for the $L = 0$ waves ($J^\pi = 0^+$, $1^+$) can be obtained by including a rather small number of channels.  This is due mainly to the Pauli principle which limits overlaps between the four nucleons.  As a consequence, the internal part is rather small and does not require large number of channels to be well described.

On the other hand, for $L = 1$ waves ($J^\pi = 0^-$, $1^-$ and $2^-$) the convergence rate is slow and many channels have to be included.  In these cases, the interaction between the $p$ and \nuc{3}{He} clusters is very attractive (it has been speculated that some $4N$ resonant states exist) and the construction of the internal wave function is more delicate.

Finally, the contribution from $L = 2$ waves is rather tiny, since the centrifugal barrier does not allow the two clusters to come close, and the corresponding phase-shifts can be calculated with good approximation by neglecting the internal part $\Psi^{LSJJ_z\pi}_C$.  Contributions from $L = 3$ or higher waves has been disregarded since they are assumed to be negligible.

Let us discuss in detail the convergence of the HH calculation of the $0^-$ phase-shift; the other $J^\pi$ states will be reported elsewhere.  As shown in Section \ref{sec:theory}, for this state $L, S = 1, 1$ and ${\mathcal S}^0_{11,11}$ can be parametrized as $\eta \exp(2{\rm i} \delta)$.  The results obtained for $\eta$ and $\delta$ at $\Ep = 4.05$  are reported in Table~\ref{table:conv}.  Here we have considered the AV18 potential model~\cite{wir95}; however, the electromagnetic interaction has been limited to just the point-Coulomb potential.  We have used $1/M_N = 41.47108$ MeV fm$^2$.  We study the convergence as explained in Section~\ref{sec:theory}, and the results presented in Table~\ref{table:conv} are arranged accordingly.  For example, the phase-shift $\delta$ reported in a row with a given set of values of $K_1, \ldots, K_7$ has been obtained by including in the expansion all the HH functions of class ${\rm C}i$ with $K\le K_i$, $i = 1, \ldots, 7$.

\begin{table}\centering
\begin{ruledtabular}

	\caption[Table]{\label{table:conv} Convergence of $0^-$ inelasticity parameter $\eta$ and phase-shift $\delta$ at $\Ep = 4.05$ MeV corresponding to the inclusion in the internal part of the wave function of the different classes C1~-~C7 in which the HH basis has been subdivided. The AV18 potential is considered here with the inclusion of the point Coulomb interaction.}

\begin{tabular}{c@{$\quad$}c@{$\quad$}c@{$\quad$}c@{$\quad$}c@{$\quad$}c@{$\quad$}c@{$\qquad$}@{$\quad$}c@{$\qquad$}c@{$\ $}}
$K_1$ & $K_2$ & $K_3$ & $K_4$ & $K_5$ & $K_6$ & $K_7$ & $\eta$ & $\delta$\ (deg) \\
\hline
 21 &&&&&&& 1.00032  & 10.649     \\
 31 &&&&&&& 1.00069  & 11.484     \\
 41 &&&&&&& 1.00107  & 11.882     \\
 51 &&&&&&& 1.00133  & 12.060     \\
 61 &&&&&&& 1.00146  & 12.136     \\
\hline
 61 &11&&&&&& 1.00139  &  12.599   \\
 61 &21&&&&&& 1.00131  &  12.897   \\
 61 &31&&&&&& 1.00134  &  13.020   \\
 61 &37&&&&&& 1.00136  &  13.055   \\
\hline
 61 &37&11&&&&& 1.00064  &  15.284   \\
 61 &37&21&&&&& 1.00049  &  15.923   \\
 61 &37&31&&&&& 1.00048  &  16.105   \\
 61 &37&35&&&&& 1.00048  &  16.132   \\
\hline
 61 &37&35&11&&&&  1.00045 & 16.256 \\
 61 &37&35&21&&&&  1.00040 & 16.646  \\
 61 &37&35&25&&&&  1.00040 & 16.727  \\
 61 &37&35&31&&&&  1.00040 & 16.794  \\
\hline
 61& 37& 35& 31&  3&  &  &  1.00012&  16.877\\
 61& 37& 35& 31&  7&  &  &  1.00002&  17.003\\
 61& 37& 35& 31& 11&  &  &  1.00000&  17.101\\
 61& 37& 35& 31& 15&  &  &  1.00000&  17.157\\
 61& 37& 35& 31& 19&  &  &  1.00000&  17.191\\
\hline
 61& 37& 35& 31&  19& 11 &  &  1.00000&  17.194\\
\hline
 61& 37& 35& 31&  19& 11 & 11 &  1.00000&  17.219\\

\end{tabular}
\end{ruledtabular}
\end{table}

The convergence of the class C1 is rather slow and a fairly large value of $K$ has to be used.  The inclusion of the second and third classes increases the phase-shift by about $4\degree$.  The class C4 contributes for additional $0.6\degree$. The number of the states of this class increases very rapidly with $K_4$ but fortunately the convergence is reached around $K = 21$. Up to now, the expansion includes only states with $\Sigma = 1$.  The contribution of the states with $\Sigma = 2$, first appearing when the class C5 is considered, is rather small, and the contribution of the states with $\Sigma = 0$ (class C6) is practically negligible.  The contribution of the class C7 is also small.  Since the number of states of this class is very large (there are $121$ channels with $\ell_1 + \ell_2 + \ell_3 = 5$ for $J^\pi = 0^-$) when confronted with a very tiny change of the phase-shift, a selection of the states has to be performed to save computing time and to avoid loss of numerical precision.  In the present example, only the channels with $(\ell_1,\ell_2,\ell_3)=(1,2,2)$ have been found important.  Note that at lower energies the convergence is noticeably faster (see below).

The convergence rate when considering the AV18/UIX interaction model is similar to the AV18 case. Since the models most frequently used for the $3N$ interactions lack a strongly-repulsive core at short interparticle distances, the convergence rate of the various classes is found not to change appreciably.

\begin{figure}
	\includegraphics[width=\columnwidth]{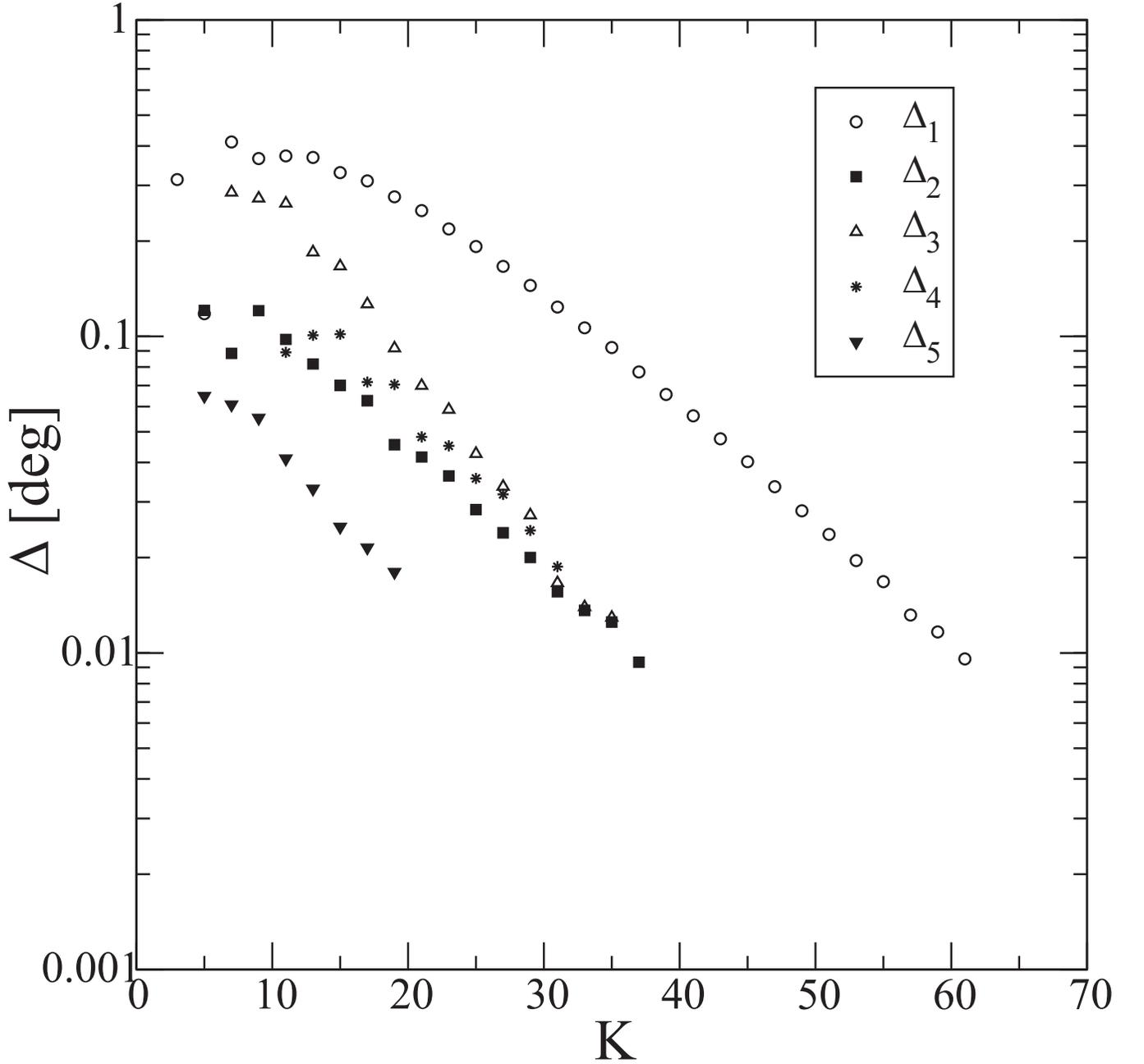}

	\caption{$0^-$ phase-shift differences for \phe elastic scattering at $\Ep = 4.05$ MeV for the classes C1 (circles), C2 (squares), C3 (up triangles) C4 (asterisks), and C5 (down triangles) as function of the grand angular value $K$ (see the text for more details). The potential used is AV18.}

	\label{fig:diff}
\end{figure}

In order to obtain a quantitative estimate of the ``missing'' phase-shift due to the truncation of the HH expansion of the various classes, let us consider $\delta(K_1, K_2, K_3, K_4, K_5, K_6, K_7)$, the phase-shift obtained by including in the expansion all the HH states of the class C1 with $K \le K_1$, all the HH states of the class C2 having $K \le K_2$, etc. Let us compute
\begin{eqnarray}
  \Delta_{1}(K)&=&\delta(K,0,0,0,0,0,0)-\delta(K-2,0,0,0,0,0,0)\ ,
  \label{eq:c1diffa}\\
  \Delta_{2}(K)&=&\delta(\overline{K_{1}},K,0,0,0,0,0)-
                  \delta(\overline{K_{1}},K-2,0,0,0,0,0)\ ,
                 \qquad \overline{K_{1}}=61\ ,
  \label{eq:c2diff}\\
  \Delta_{3}(K)&=&\delta(\overline{K_{1}},\overline{K_{2}},K,0,0,0,0)-
                  \delta(\overline{K_{1}},\overline{K_{2}},K-2,0,0,0,0)\ ,
                 \qquad \overline{K_{2}}=37\ ,
  \label{eq:c3diff}
\end{eqnarray}
and so on.  The values obtained for $\Delta_{i}$, $i=1$, $5$ are shown in Fig.~\ref{fig:diff} for the $\Ep = 4.05$ MeV on a logarithmic scale.  As can be seen, all the differences $\Delta_1$ through $\Delta_5$ decrease exponentially, and approximately with the same decay constant.  For a given $K$, however, there is a clear hierarchy $\Delta_1(K) \gg \Delta_2(K) \approx \Delta_3(K) \approx \Delta_4(K) \gg \Delta_5(K)$. Note that there are slight fluctuations in $\Delta(K)$ as $K$ is increased (this is evident particularly for $\Delta_5$). The phase-shift differences for the classes C6 and C7 are not shown since they are tiny.

From the simple behavior observed in Fig.~\ref{fig:diff}, we can readily estimate the missing phase-shift due to the truncation of the expansion to finite values of $K = \overline{K}$.  Let us suppose that the states of class $i$ up to $K = \overline{K}$ have been included and used to compute $\Delta_i(\overline{K})$.  From the observed behavior $\Delta_i(K) \propto \exp(-\gamma K)$, the ``missing'' phase-shift $\delta^M_i$ due to the states with $K = \overline{K}+2$, $\overline{K}+4$, $\ldots$, can be estimated as
\begin{equation}
	\delta^M_i =  c(\gamma) \; \Delta_i(\overline{ K})\ ,
	\label{eq:extra}
\end{equation}
where
\begin{equation} \nonumber
  c(\gamma)= \sum_{K=\overline{ K}+2,\overline{K}+4,\ldots}^\infty  e^{-\gamma(K-\overline{ K})} = {\frac{x}{1-x}} \ ,
  \qquad  \mbox{and} \qquad
  x=e^{-2\gamma} \ .
\end{equation}
For example, consider the missing phase shift for the class C1.  For $\overline{ K} = 61$, $\Delta_1(\overline{ K}) = 0.009\degree$ and $x \approx 0.8$.  Therefore, $\delta^M_1 \approx 0.04\degree$, a rather small quantity.  The missing phase-shift of the other classes can be estimated in the same way.

However, to estimate the total missing phase-shift $\delta^M_T$ due to the truncation of the expansion of the first class up to $K\le K_1$, of the second class up to $K\le K_2$, etc., we cannot simply add the $\Delta^M_i$, $i=1,\ldots,7$ so obtained.  The inclusion of the HH states of classes C2, C3, $\ldots$, also alters the convergence of class C1 by a small amount, etc.  To study the ``full'' rate of convergence, we have taken advantage of the fact that the various $\Delta(K)$ show a similar convergence behavior (with approximately the same decay constant $\gamma$) and that the coefficient $c(\gamma)$ defined in Eq.~(\ref{eq:extra}) does not depend on $\overline{ K}$. Let us then consider
\begin{equation}
  \Delta_T(K_1,\ldots,K_7)= \delta(K_1,\ldots,K_7)-\delta(K_1-2,\ldots,K_7-2)
  \ .\label{eq:dbet}
\end{equation}
From the above discussion, we can estimate the total missing phase shift as
\begin{equation}
   \delta^M_T =   {\frac{x}{1-x}} \; \Delta_T(\overline{K}_1,\ldots,\overline{K}_7)\ ,
    \qquad x=e^{-2\gamma} \ .
   \label{eq:extra2}
\end{equation}
As an example, the values for $\Delta_i(K_1,\ldots,K_7)$ computed at $\Ep = 1.00$, $2.25$, and $4.05$ MeV are reported in Table~\ref{table:extra}, from which it is possible to derive the values of $\gamma$ and then of $\delta^M_T$.  The computed values of  $\delta^M_T$ using Eq.~\ref{eq:extra2} are reported at the bottom.

\begin{table*}
	\caption{\label{table:extra} Convergence of $0^-$ phase-shift at $Ep = 1.00$, $2.25$ and $4.05$ MeV  corresponding to the inclusion in the internal part of the wave function of the different subsets of HH basis.  The AV18 potential is considered here with the inclusion of the point Coulomb interaction.  In the last row, the total missing phase-shifts computed as described in text have been reported (we have estimated $e^{-2\gamma}\approx0.8$).}

\begin{ruledtabular}
\begin{tabular}{c@{$\quad$}c@{$\quad$}c@{$\quad$}c@{$\quad$}
                c@{$\quad$}c@{$\quad$}c@{$\qquad$}@{$\quad$}
                c@{$\qquad$}c@{$\qquad$}c@{$\ $}}
 && & & & & & $\Ep=1.00$\ MeV& $\Ep=2.25$\ MeV& $\Ep=4.05$\ MeV\\
$K_1$ & $K_2$ & $K_3$ & $K_4$ & $K_5$ & $K_6$ & $K_7$ & 
     $\delta$\ (deg) & $\delta$\ (deg) & $\delta$\ (deg)\\
\hline
 51& 27& 25& 21&  1 & 1 & 1  &  1.867  & 7.401 & 16.364 \\
 53& 29& 27& 23&  3 & 3 & 3  &  1.880  & 7.467 & 16.559 \\
 55& 31& 29& 25&  5 & 5 & 5  &  1.891  & 7.526 & 16.724 \\
 57& 33& 31& 27&  7 & 7 & 7  &  1.902  & 7.583 & 16.878 \\
 59& 35& 33& 29&  9 & 9 & 9  &  1.912  & 7.632 & 17.010 \\
 61& 37& 35& 31& 11 & 11& 11 &  1.919  & 7.668 & 17.106 \\
\hline
   &   &   & $\delta^M_T$ &    &   &    &  0.028  & 0.144 &  0.384 \\
\end{tabular}
\end{ruledtabular}
\end{table*}

As can be seen in Table \ref{table:extra}, $\delta^M_T$ is estimated to be rather small at $\Ep = 1.00$ and $2.25$ MeV. However, for the largest energy the convergence seems not to be completely under control and higher values of $K_1$ through $K_7$ should be employed.  In any case, we can see that the missing phase-shift is less than 2 \%. Analogous problems have been found for the $1^-$ and $2^-$ states, whereas for the other states the convergence did not present any difficulty.

\begin{figure}
	\includegraphics[width=5in]{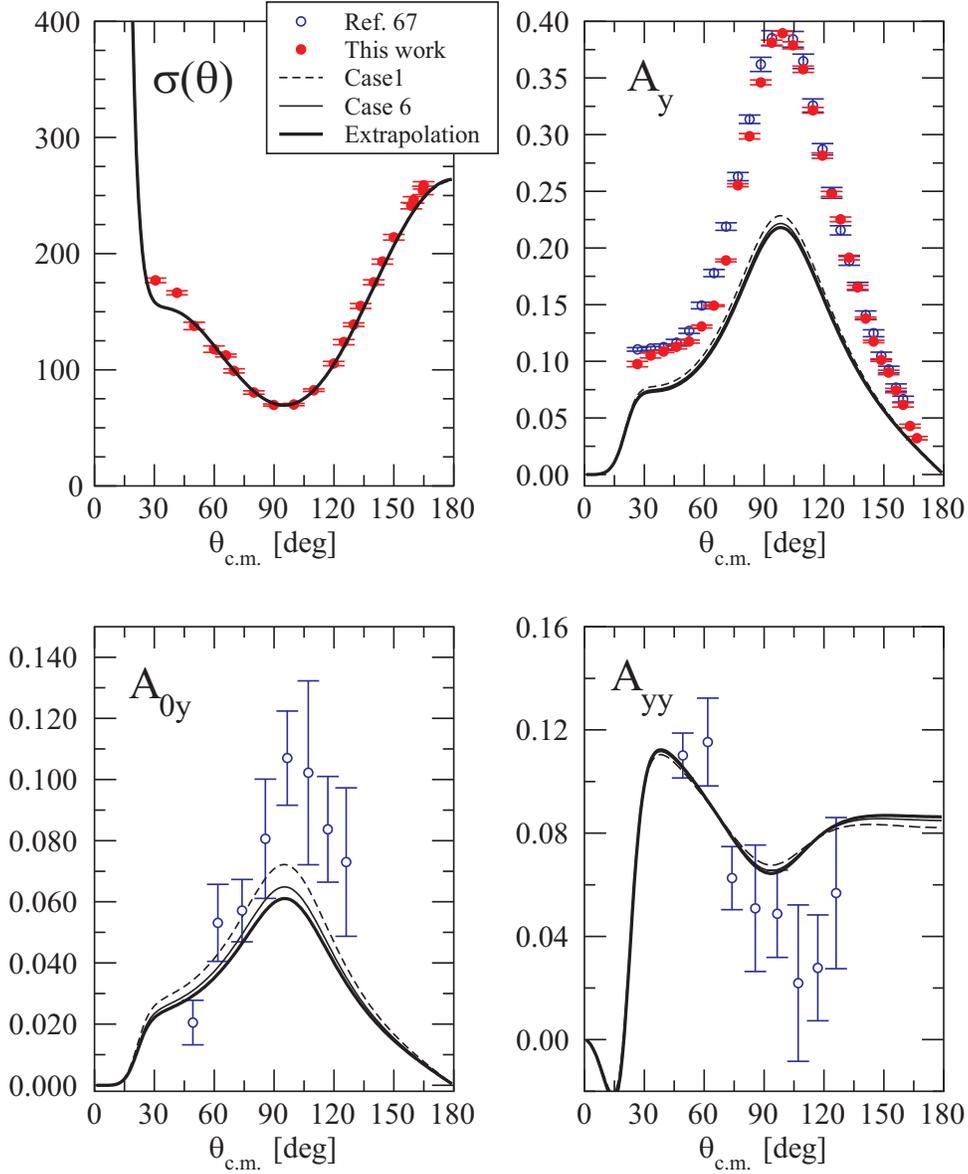}

	\caption{(Color online) Four \phe elastic scattering observables at $\Ep = 4.05$ MeV calculated using different values for the $0^-$ phase-shift. The dashed and solid curves were obtained using different truncations of the HH expansion corresponding to choices of $(K_1,\ldots,K_7)$ reported in the first and sixth rows of Table~\protect\ref{table:extra}. The thick solid curves have been obtained using the extrapolated value for this phase-shift, obtained as explained in the text. The experimental data are from Ref.~\protect\cite{all93a} (open circles) and from the present work (solid circles). The potential used is AV18.}

	\label{fig:conv}
\end{figure}

Since at the moment the inclusion of a greater number of states would require a significant increase in computing time, we have preferred to use the extrapolation outlined above for obtaining estimates for the converged phase-shifts and mixing parameters for the $J^\pi = 0^-$, $1^-$ and $2^-$ states.  To show the effect of the extrapolation on the observables, we present in Fig.~\ref{fig:conv} the results for four \phe observables at $\Ep = 4.05$ MeV and calculations using the AV18 potential. The dashed and thin solid curves have been obtained using the $0^-$ phase shift calculated with different values for $K_1, \ldots, K_7$. More precisely, the dashed (solid) curves have been obtained using the value $\delta = 16.364\degree$ ($17.106\degree$) obtained with the choice of $K_1, \ldots, K_7$ reported in the first (sixth) row of Table~\ref{table:extra}. The thick solid curve has been obtained using the extrapolated value $\delta(61, 37, 35, 31, 11, 11, 11) +\delta^M_T \approx 17.5\degree$. The other phase-shift were taken with their final values (in particular, for the $1^-$ and $2^-$ phase-shifts, we used the extrapolated values obtained using a similar procedure as described above).  The four observables considered in the Figure are: the differential cross section \dsig, the proton analyzing power \ay, the \nuc{3}{He} analyzing power \ayhe, and the spin correlation coefficient \ayy.

As can be seen, there is a good convergence for \dsig, \ay, and \ayy. The observable \ayhe is more sensitive to $\delta$, and the convergence is more critical. In any case, however, the thick solid curves are rather close to the thin solid curves, showing that the convergence has been nearly reached. A similar analysis has been performed also for the $1^-$ and $2^-$ phase-shifts and mixing parameters, with similar findings. Therefore, we can conclude that the convergence  of the HH expansion is sufficiently good to obtain nearly correct predictions for the \phe observables and allowing therefore for meaningful comparisons between theory and experimental data.  The calculated phase shifts are in reasonable agreement with those calculated by other methods (for example, those in Ref. \onlinecite{rei03}).

\begin{acknowledgments}
The authors would like to thank W. Tornow, A. Fonseca, and T. B. Clegg for very useful discussions;  T. V. Daniels and M. S. Boswell for their assistance in the data taking; M. H. Wood for assistance with the gas-jet target; and J. D. Dunham, E. P. Carter, and especially R. M. O'Quinn for copious technical assistance.  This work was supported in part by the U.S. Department of Energy under Grant No. DE-FG02-97ER41041.
\end{acknowledgments}


\end{document}